\documentclass[a4paper,11pt]{article}
\pdfoutput=1 

\usepackage{jheppub} 
\usepackage[T1]{fontenc} 
\usepackage[all]{hypcap} 
\usepackage{slashed} 
\usepackage{multirow} 
\usepackage{booktabs} 
\usepackage{pifont} 
\usepackage{parskip} 
\usepackage[section]{placeins} 

\graphicspath{ {../plots/} {plots/} }

\newcommand{\cmark}{\ding{51}}
\newcommand{\xmark}{\ding{55}}
\newcommand{\subh}{\textit{\fontsize{1.5mm}{0mm}H}}

\newcommand{\limitHiggs}{607}
\newcommand{\limitEPWT}{405}
\newcommand{\limitHiggsEWPT}{694}
\newcommand{\limitHiggsEWPTCaseB}{560}
\newcommand{\limitDirectSearches}{638}

\title{\boldmath Littlest Higgs with T-parity: Status and Prospects}
\author{J\"urgen Reuter,}
\author{Marco Tonini,}
\author{Maikel de Vries}
\affiliation{DESY Theory Group\\Notkestr. 85, 22607 Hamburg, Germany}
\emailAdd{juergen.reuter@desy.de}
\emailAdd{marco.tonini@desy.de}
\emailAdd{maikel.devries@desy.de}
\preprint{\small DESY 13-123}
\arxivnumber{1310.2918}

\abstract{
The Littlest Higgs model with T-parity is providing an attractive solution to the fine-tuning problem. This solution is only entirely natural if its intrinsic symmetry breaking scale $f$ is relatively close to the electroweak scale. We examine the constraints using the latest results from the $8 \, \textrm{TeV}$ run at the LHC. Both direct searches and Higgs precision physics are taken into account. The constraints from Higgs couplings are by now competing with electroweak precision tests and both combined exclude $f$ up to $\limitHiggsEWPT \, \textrm{GeV}$ or $\limitHiggsEWPTCaseB \, \textrm{GeV}$ depending on the implementation of the down-type Yukawa sector. Direct searches provide robust and complementary limits and constrain $f$ to be larger than $\limitDirectSearches \, \textrm{GeV}$. We show that the Littlest Higgs model parameter space is slowly driven into the TeV range. Furthermore, we develop a strategy on how to optimise present supersymmetry searches for the considered model, with the goal to improve the constraints and yield more stringent limits on $f$.
}

\begin{document}

\maketitle
\flushbottom

\newpage 

\section{Introduction}
\label{sec:intro}
Since the discovery of a bosonic resonance at $126 \, \textrm{GeV}$ \cite{Chatrchyan:2012ufa,Aad:2012tfa} the fine-tuning problem of the Higgs boson has become even more intriguing. Moreover, we are still lacking a true microscopic picture of electroweak symmetry breaking. In the past various models have been proposed to regulate quadratically divergent contributions to the Higgs mass. In the regime of weakly coupled physics, supersymmetry is the most promising candidate, and at the LHC various searches have been performed. In the strongly coupled regime, Composite Higgs is the most adopted candidate, among them the class of Little Higgs models \cite{ArkaniHamed:2001nc,ArkaniHamed:2002pa,ArkaniHamed:2002qy,Low:2002ws,Kaplan:2003uc,Schmaltz:2004de,Cheng:2003ju,Cheng:2004yc,Pappadopulo:2010jx,Schmaltz:2010ac,Martin:2013fta,Kearney:2013cca,Kilian:2004pp,Kilian:2006eh}. The focus of the experimental collaborations up to now has been on supersymmetry and therefore exclusion limits for other classes of models are either indirect or simply not known. In order to gain as much discriminative power out of LHC data, it is therefore very important to undertake the endeavour and try to constrain these kinds of models in a manner as exhaustive as is done for supersymmetry.

Little Higgs models are a special class of composite models based on a collective symmetry breaking pattern, thereby ameliorating the little hierarchy problem. For these models the key observation is that the fine-tuning is proportional to $(v/f)^2$. Here, $f = \Lambda / (4\pi)$ is the collective symmetry breaking scale, and $\Lambda$ the UV cut-off of the model. Hence if these models want to address their original purpose, the scale $f$ should not exceed the value of $1 \, \textrm{TeV}$ by too much, since that would already imply fine-tuning in the percent range. In this light it is interesting to evaluate constraints from electroweak precision physics, Higgs precision physics and direct searches for realistic Little Higgs models.

The original constructions of Little Higgs models suffered severely from electroweak precision tests (EWPT) \cite{Csaki:2002qg,Hewett:2002px,Kilian:2003xt}, which led to the introduction of a discrete symmetry called T-parity or to new model building approaches, in particular with the introduction of a second nonlinear sigma field that couples only to the gauge bosons, disconnecting the mass of the heavy gauge bosons from the mass of the top partners. Both methods found their realizations in the Littlest Higgs with T-parity \cite{ArkaniHamed:2002qy,Cheng:2003ju,Cheng:2004yc,Pappadopulo:2010jx} and in the recent Bestest Little Higgs \cite{Schmaltz:2010ac} and Next to Littlest Higgs \cite{Martin:2013fta} models, respectively. For these models EWPTs are much less severe and constrain the symmetry breaking scale e.g. only up to roughly $400 \, \textrm{GeV}$ in the T-parity scenario \cite{Reuter:2012sd,Hubisz:2005tx,Berger:2012ec,Asano:2006nr}. However now, as a great number of LHC results become available, the limits on these models have to be revisited once again. In this paper we consider the Littlest Higgs with T-parity (LHT) and provide an update for the constraints from Higgs precision physics as presented in \cite{Reuter:2012sd}. In addition, all direct searches from CMS and ATLAS will be analysed and recasted for this model whenever feasible. Finally this will result in a lower bound on the symmetry breaking scale $f$ from all possible corners of collider physics and will provide the most stringent test on the compatibility of the LHT model with experimental data. 

The structure of the paper is the following. In section \ref{sec:lhtmodel} we review the theoretical set-up of the Littlest Higgs model with T-parity as far as it is necessary to understand the constraints and limits. This section is as general as possible and details relevant for LHC phenomenology will be emphasised. Section \ref{sec:lhtphenomenology} contains a full treatment of the phenomenological details of the LHT model tailored to $8 \, \textrm{TeV}$ experimental searches, including an analysis of the experimental final states for this model which enables identification of the relevant searches by the ATLAS and CMS experiments for the $8 \, \textrm{TeV}$ run. This endeavour has been undertaken in section \ref{sec:experimentalsearches} where exclusion limits on the LHT parameter space are presented. Then we comment on optimising existing searches for the LHT model in section \ref{sec:optimisingsusy}. Finally concluding remarks and an outlook are presented in section \ref{sec:conclusion}.

\section{LHT model}
\label{sec:lhtmodel}
As already mentioned in the introduction, the discovery of a bosonic state compatible with the electroweak fit to the SM has confirmed the existence of a particle acting as the Higgs particle of the SM. So far, no  deviations of this state from its SM properties have been seen in the LHC data. Clearly, this was not expected without the observation of new physics given the tight constraints from EWPT on the Higgs boson. Almost all models beyond the SM (BSM) have the SM with a Higgs boson as low-energy effective theory. However, the SM does not provide a microscopic explanation for EWSB, that is a dynamic mechanism for EWSB itself or an explanation for the generation of its scalar potential. Furthermore, at the moment it is not clear whether the SM without further constituents or components could serve as a stable low-energy theory below the Planck scale. 

Little Higgs models overcome this weakness by giving a dynamical explanation for EWSB in the form of a condensing new matter sector underlying new strong interactions. The strong scale $\Lambda$ is supposed to lie in the region of a few tens of TeV. From this arises a weakly coupled theory at a scale $f = \Lambda / (4\pi)$ roughly at a few TeV due to a pattern of global and gauged symmetries that are intertwined in a way to generate a little hierarchy of two orders of magnitude between the strong scale $\Lambda$ and the electroweak scale $v$. This is necessary to prevent the strong dynamics sector to generate unacceptably large contributions to the electroweak precision observables. But there are still additional contributions to the EWPT, coming from the introduction of new gauge and fermionic degrees of freedom necessary to implement the aforementioned enlarged global and gauge symmetries of Little Higgs models in the SM. To remove these contributions, a new discrete symmetry, T-parity has been introduced \cite{Cheng:2003ju,Cheng:2004yc}.

There are two different classes of Little Higgs models, depending on whether the embedding of the additional gauge symmetries is done in a product group or in a simple group. The implementation of T-parity in simple group models is rather difficult, as there is always a remaining even new neutral gauge boson. Hence, in this paper we will focus on the most popular implementation of the product group set-up, namely the Littlest Higgs model with T-parity \cite{ArkaniHamed:2002qy,Cheng:2003ju,Cheng:2004yc}. We mainly follow the presentations given in \cite{Han:2003wu,Kilian:2003xt,Hubisz:2004ft,Blanke:2006eb,Belyaev:2006jh,Chen:2006cs,Hubisz:2005tx}. The model is based on a $SU(5)/SO(5)$ coset space, while the gauge group is $G \times SU(3)_c \times SU(2)^2 \times U(1)^2$, with $G$ the gauge group of the strong sector, and the doubled electroweak gauge group. 

Note that the introduction of T-parity has the additional benefit of providing a stable, weakly interacting particle by means of the lightest particle odd under T-parity. The constraints from dark matter experiments and cosmic microwave background for the LHT model are not discussed here: the latest results can be found in \cite{Asano:2006nr,Wang:2013yba}.

In this brief introduction the focus will be on details relevant for LHC collider phenomenology: we will first discuss the gauge sector, then the scalar sector, and finally the fermionic sector of the model. This collects the independent parameters of the model, and their connection to the masses of the new states as well as their couplings. Mass and coupling relations relevant for the LHT phenomenology in the following sections are highlighted in boxes.

\subsection{Gauge sector}
\label{sec:gaugesector}
The global symmetry structure of the LHT model is defined by the coset space
\begin{equation} \label{eq:lht}
	SU(5) / SO(5) ,
\end{equation}
where the spontaneous symmetry breaking is realised at the scale $f$ via the vacuum expectation value (\emph{vev}) of an $SU(5)$ symmetric tensor field
\begin{equation}
	\langle \Sigma \rangle = \begin{pmatrix} \mathbf{0}_{2 \times 2} & \mathbf{0}_{2 \times 1} & \mathbf{1}_2 \\ \mathbf{0}_{1 \times 2} & 1 & \mathbf{0}_{1 \times 2} \\ \mathbf{1}_2 & \mathbf{0}_{2 \times 1} & \mathbf{0}_{2 \times 2} \end{pmatrix} .
\end{equation}
Fourteen Nambu-Goldstone Bosons (NGBs) $\Pi^a$ with $a = 1, \ldots, 14$ arise in this set-up, parametrised in the usual nonlinear sigma model formalism as
\begin{equation}
	\Sigma(x) = e^{2 \, i \, \Pi^a X^a(x)/f} \, \langle \Sigma \rangle \equiv \xi^2 (x) \langle \Sigma \rangle ,
\end{equation}
where $X^a$ are the broken generators of the coset space \eqref{eq:lht}. 

As mentioned above, this model belongs to the class of product group models, where the SM gauge group emerges from the diagonal breaking of the product of several gauged groups: there is indeed a local invariance under $[SU(2)_{1} \otimes U(1)_{1}] \otimes [SU(2)_{2} \otimes U(1)_{2}]$, embedded in the matrix structure, spontaneously broken via $\langle \Sigma \rangle$ to its diagonal subgroup, which is identified with the SM gauge group. Explicitly, the kinetic term for the NGB matrix can be expressed in the standard nonlinear sigma model formalism as 
\begin{equation} \label{eq:scalkin}
	\mathcal{L}_{\Sigma} = \frac{f^2}{8} \textrm{Tr} \big| D_\mu \Sigma \big|^2 ,
\end{equation}
where the covariant derivative is defined as 
\begin{equation}
	D_\mu \Sigma = \partial_\mu \Sigma - i \sum_{j=1}^2 \left[ g_j \left( W_j \Sigma + \Sigma W_j^t \right) + g_j^\prime \left( B_j \Sigma + \Sigma B_j^t \right) \right] .
\end{equation}
The generators of the gauged symmetries are explicitly given as 
\begin{align}
	Q_1^a = & \begin{pmatrix} \sigma^a / 2 & 0 & 0 \\ 0 & 0 & 0 \\ 0 & 0 & 0 \end{pmatrix} & Y_1 = \tfrac{1}{10} \textrm{diag} \left( 3, 3, -2, -2, -2 \right) \nonumber \\ 
	Q_2^a = & \begin{pmatrix} 0 & 0 & 0 \\ 0 & 0 & 0 \\ 0 & 0 & -\sigma^{a \ast} / 2 \end{pmatrix} & Y_2 = \tfrac{1}{10} \textrm{diag} \left( 2, 2, 2, -3, -3 \right) . 
\end{align}

In the gauge boson sector, T-parity is introduced as an exchange symmetry between the gauge bosons of the two different copies of the SM gauge group as 
\begin{equation} 	\label{eq:tparity}
	T: \qquad W_{1 \, \mu}^a \leftrightarrow W_{2 \, \mu}^a, \qquad B_{1 \, \mu} \leftrightarrow B_{2 \, \mu} .
\end{equation}
This is originally inherited from the corresponding transformation properties of the Lie algebra generators \cite{Cheng:2003ju,Cheng:2004yc}. The gauge-kinetic Lagrangian \eqref{eq:scalkin} of the Littlest Higgs model is then invariant under T-parity for 
\begin{equation}
	g_1 = g_2 = \sqrt{2} g, \qquad g_1^\prime = g_2^\prime = \sqrt{2} g^\prime .
\end{equation}
A set of $SU(2) \otimes U(1)$ gauge bosons ($W^{a \, \prime}$, $B^\prime$) obtains a mass term of order $f$ from \eqref{eq:scalkin}, while the other set ($W^a$, $B$) remains massless and is identified with the SM gauge bosons. The mass eigenstates are related to the gauge eigenstates by the following field rotations 
\begin{align}
	W^a = & \frac{1}{\sqrt{2}} \left( W^a_1 + W^a_2 \right) \qquad B = \frac{1}{\sqrt{2}} \left( B_1 + B_2 	\right) \nonumber \\ 
	W^{a \, \prime} = & \frac{1}{\sqrt{2}} \left( W^a_1 - W^a_2 \right) \qquad B^{\prime} = \frac{1}{\sqrt{2}} 	\left( B_1 - B_2 \right) .
\end{align}
Clearly, under T-parity \eqref{eq:tparity} the heavy gauge bosons are odd while the SM ones are even.

EWSB induces further mixing for the light and heavy gauge bosons separately: in particular the mass eigenstates in the neutral heavy sector will be a linear combination of $W^{3 \, \prime}$ and $B^{\prime}$, producing a heavy partner of the photon $A_\subh$ and of the $Z$ boson $Z_\subh$, with a mixing angle of the order of $v^2 / f^2$
\begin{equation}
	\sin \theta_\subh \simeq \frac{5 g g^\prime}{4 \left( 5 g^2-g^{\prime \, 2} \right)} \frac{v^2}{f^2} .
\end{equation}
In here $v$ represents the \emph{vev} of the Higgs doublet, whose dynamical generation will be described later in the text. At $\mathcal{O} \left( v^2 / f^2 \right)$ in the expansion of the Lagrangian \eqref{eq:scalkin}, the mass spectrum after EWSB is given by
\begin{align}
	& m_W = \frac{g v}{2} \left( 1 - \frac{1}{12}\frac{v^2}{f^2} \right) \qquad m_Z = \frac{g v}{2 c_w} \left( 1 - \frac{1}{12}\frac{v^2}{f^2} \right)  \qquad m_\gamma = 0 \nonumber \\
	& \boxed{m_{W_\subh} = m_{Z_\subh} = g f \left( 1- \frac{1}{8} \frac{v^2}{f^2} \right)} \qquad \boxed{m_{A_\subh} = \frac{g^\prime f}{\sqrt{5}} \left( 1- \frac{5}{8} \frac{v^2}{f^2} \right)} \, .
\end{align}
In order to match the Standard Model prediction for the gauge boson masses, the \emph{vev} needs to be redefined in terms of the typical SM value $v_\textrm{SM} = 246 \, \textrm{GeV}$ via the functional form
\begin{equation}
	v = \frac{f}{\sqrt{2}} \arccos{\left( 1 - \frac{v_\textrm{SM}^2}{f^2} \right)} \simeq v_\textrm{SM} \left( 1 + \frac{1}{12} \frac{v_\textrm{SM}^2}{f^2} \right) .
\end{equation}

\subsection{Scalar sector}
\label{sec:scalarsector}
Under the unbroken $SU(2)_L \otimes U(1)_Y$ the Goldstone bosons transform as $\textbf{1}_0 \oplus \textbf{3}_0 \oplus \textbf{2}_{1/2} \oplus \textbf{3}_{\pm 1}$. The $\textbf{2}_{1/2}$ component is identified with the Higgs doublet $H$, while the $\textbf{3}_{\pm 1}$ component is a complex triplet under $SU(2)_L$ which forms a symmetric tensor
\begin{equation}
	\Phi = \frac{- i}{\sqrt{2}} \begin{pmatrix} \sqrt{2} \phi^{++} & \phi^+ \\ \phi^+ & \phi^0 + i \, \phi^P \end{pmatrix} .
\end{equation}
Both $\phi^0$ and the pseudoscalar $\phi^P$ are real scalars, whereas the $\phi^{++}$ and $\phi^+$ are complex scalars. The other Goldstone bosons are the longitudinal modes of the heavy gauge bosons and therefore will not appear in unitary gauge. In the latter gauge, the Goldstone boson matrix $\Pi$ is given by
\begin{equation}
	\Pi = \frac{1}{\sqrt{2}} \begin{pmatrix} 0 & H & \sqrt{2} \Phi \\ H^\dagger & 0 & H^t \\ \sqrt{2} \Phi^\dagger & H^\ast & 0 \end{pmatrix} .
\end{equation}
The action for T-parity in the scalar sector is defined as
\begin{equation} \label{eq:sigmat}
	T: \quad \Pi \rightarrow - \Omega \, \Pi \, \Omega ,
\end{equation}
where $\Omega = \textrm{diag} (1, 1, -1, 1, 1)$ is introduced to give the Higgs positive parity while keeping the triplet odd.

The global symmetries prevent the appearance of a potential for the scalar fields at tree level. The scalar potential is indeed generated dynamically at one-loop and higher orders due to the interactions with gauge bosons and fermions, and is parametrised through the Coleman-Weinberg (CW) potential \cite{Coleman:1973jx}. The most general scalar potential invariant under the SM gauge groups, involving one doublet field $H$ and one triplet field $\phi$ can be written up to dimension-four operators as
\begin{align}
	V_\textrm{CW} = \; & \lambda_{\phi^2} f^2 \textrm{Tr} \left( \phi^\dagger \phi \right) + i \lambda_{h \phi h} f \left( H \phi^\dagger H^t - H^\ast \phi H^\dagger \right) - \mu^2 H H^\dagger  \nonumber \\ 
	& + \lambda_{h^4} \left( H H^\dagger \right)^2 + \lambda_{h \phi \phi h} \, H \phi^\dagger \phi H^\dagger + \lambda_{h^2 \phi^2} \, H H^\dagger \textrm{Tr} \left( \phi^\dagger \phi \right) \nonumber \\ 
	& + \lambda_{\phi^2 \phi^2} \left[ \textrm{Tr} \left( \phi^\dagger \phi \right) \right]^2 + \lambda_{\phi^4} \, \textrm{Tr} \left( \phi^\dagger \phi \phi^\dagger \phi \right) .
\end{align}
The coefficients $\mu^2$, $\lambda_{h^2 \phi^2}$ and $\lambda_{h \phi h}$ get no contribution from the quadratically divergent part of the one-loop CW potential, either because of the collective symmetry breaking mechanism ($\mu^2$, $\lambda_{h^2 \phi^2}$) or because of T-parity ($\lambda_{h \phi h}$). They thus receive only log-divergent contributions at one-loop, and quadratically divergent contributions starting from the two-loop level. The latter suppression of $\mu^2$ from an extra loop factor gives the natural hierarchy between the electroweak scale and the cut-off scale $\Lambda$: two- and higher-loop contributions have not been calculated, and therefore $\mu^2$ can be treated as a free parameter. Its value will be fixed by the observed Higgs mass \eqref{eq:scalars}. Since the quartics $\lambda_{h^2 \phi^2}$, $\lambda_{h \phi h}$ are two-loop suppressed as well, they are negligible with respect to the other $\mathcal{O}(1)$ quartic couplings, and therefore we will not consider them.

The remaining coefficients can be expressed in terms of the fundamental parameters of the model
\begin{align}
	& \lambda_{\phi^2} = 2 (g^2 + g^{\prime \, 2}) + 8 \lambda_1^2 & \lambda_{h^4} = \frac{1}{4} \lambda_{\phi^2} \nonumber \\
	& \lambda_{h \phi \phi h} = -\frac{4}{3} \lambda_{\phi^2} & \lambda_{h^2 \phi^2} = -16 \, \lambda_1^2 \nonumber \\
	& \lambda_{\phi^4} = -\frac{8}{3} (g^2 + g^{\prime \, 2}) + \frac{16}{3} \lambda_1^2 , &
\end{align} 
where $\lambda_1 = \lambda_1 (R, m_t)$ is a parameter of the third generation fermion sector which will be explained in the next subsection. Minimising the potential to obtain the doublet \emph{vev} $v$ which triggers EWSB, one can express the parameters in the scalar potential in terms of the physical parameters $f$, $m_h$ and $v$. Diagonalising the scalar mass matrix, one obtains the following spectrum
\begin{equation} \label{eq:scalars}
	m_h = \sqrt{2} \mu \qquad \boxed{m_{\Phi} = \frac{\sqrt{2} m_h}{v} f } \, ,
\end{equation}
where all components of the triplet $\left( \phi^{++}, \phi^+, \phi^0, \phi^P \right)$ are degenerate at the order we are considering. Since $\mu^2$ is treated as a free parameter, we will assume the measured Higgs mass for the scalar doublet $h$, fixing therefore the value of $\mu$.  

\subsection{Fermion sector}
\label{sec:fermionsector}
To implement T-parity in the fermion sector one introduces two $SU(2)_A$ fermion doublets $q_A =\left( i d_{L_A},-i u_{L_A} \right)^t$ with $A=1,2$. T-parity will then be defined such that 
\begin{equation}
	T: \quad q_1 \leftrightarrow - q_2 .
\end{equation}
The T-even combination $u_{L+} = \left( u_{L_1}-u_{L_2} \right)/\sqrt{2}$ will be the up-type component of the SM fermion doublet, while the T-odd combination $u_{L-} = \left( u_{L_1}+u_{L_2} \right)/\sqrt{2}$ will be its T-odd partner. The same definitions hold also for the down-type components. We require that the T-even (SM) eigenstates obtain a mass only from Yukawa-like interactions after EWSB, while forcing the masses of the T-odd eigenstates to be at the TeV scale. The standard procedure is to embed the $q_A$ doublets into incomplete $SU(5)$ multiplets $\Psi_A$ as $\Psi_1 = \left( q_1, 0, \mathbf{0}_{1 \times 2} \right)^t$ and $\Psi_2 = \left( \mathbf{0}_{1 \times 2}, 0, q_2 \right)^t$, with the following transformation rules
\begin{align}
	\label{eq:quarklht}
	SU(5): \qquad & \Psi_1 \rightarrow V^\ast \Psi_1, \quad \Psi_2 \rightarrow V \Psi_2, \quad V \in SU(5) \nonumber \\
	T: \qquad & \Psi_1 \leftrightarrow - \langle \Sigma \rangle \Psi_2 .
\end{align}
To give masses to the additional fermions, an $SO(5)$ multiplet $\Psi_c$ is also introduced as $\Psi_c = \left( q_c, \chi_c, \tilde{q}_c  \right)^t$, nonlinearly transforming under the full $SU(5)$
\begin{align}
	SU(5): \qquad & \Psi_c \rightarrow U \Psi_c \nonumber \\
	T: \qquad & \Psi_c \rightarrow - \Psi_c ,
\end{align}
where the matrix $U$ is a nonlinear transformation. The components of the latter $\Psi_c$ multiplet are the so-called \emph{mirror} fermions. 

A possible $SU(5)$- and T-invariant Lagrangian that could generate a TeV scale mass only for the T-odd combinations is finally given by
\begin{align} \label{eq:Todd}
	\mathcal{L}_k = & -k f \left( \bar{\Psi}_2 \xi \Psi_c + \bar{\Psi}_1 \langle \Sigma \rangle \Omega 	\xi^\dagger \Omega \Psi_c \right) - m_q \, \bar{u}^\prime_c \, u_c - m_q \, \bar{d}^\prime_c \, d_c - m_\chi \, \bar{\chi}^\prime_c \, \chi_c \textrm{ + h.c.} \nonumber \\
	\supset & - \sqrt{2} k f \left[ \bar{d}_{L-} \, \tilde{d_c} + \frac{1 + c_\xi}{2} \, \bar{u}_{L-} \, \tilde{u}_c - \frac{s_\xi}{\sqrt{2}} \, \bar{u}_{L-} \, \chi_c - \frac{1 - c_\xi}{2} \, \bar{u}_{L-} \, u_c \right] + \nonumber \\  
	& - m_q \, \bar{u}^\prime_c \, u_c - m_q \, \bar{d}^\prime_c \, d_c - m_\chi \, \bar{\chi}^\prime_c \, \chi_c \textrm{ + h.c.} 
\end{align}
where $c_\xi = \cos \left( h / \sqrt{2} f \right)$, $s_\xi = \sin \left( h / \sqrt{2} f \right)$. Indeed, no mass term for the T-even combinations ($u_{L+}, d_{L+}$) is generated. The coupling $k$ is in general a matrix in flavour space for both quarks and leptons. As first noted in \cite{Hubisz:2005bd}, in analogy with the CKM transformations, the matrix $k^i_{\phantom{i}j}$ is diagonalised by two $U(3)$ matrices
\begin{equation}
	k^i_{\phantom{i}j} = \left( V_\subh \right)^i_k \left( k_D \right)^k_l \left( U_\subh \right)^l_j .
\end{equation}
The matrix $V_\subh$ acts on the left handed fields while $U_\subh$ acts on the right-handed $\Psi_c$ fields. The gauge interactions in the T-parity eigenbasis are given qualitatively by
\begin{equation}
	g \bar{Q}_{-i} \slashed{A}_- Q^i_+ + g \bar{Q}_{+i} \slashed{A}_- Q^i_- ,
\end{equation}
where the $A_-$ and $Q_-$ are the T-odd gauge bosons and fermions respectively, while the $Q_+$ are the T-even fermions. Rotating to the mass eigenbasis, using $H$ and $L$ indices for heavy and light, these interactions can be re-expressed as
\begin{equation}
	g \bar{Q}_{Hi} V^{\dagger i}_{Hj} \slashed{A}_\subh \begin{pmatrix} \left( V_u \right)^j_k u_L^k \\ \left( V_d \right)^j_k d_L^k \end{pmatrix} 
	+ g \begin{pmatrix} \bar{u}_{Lk} ( V_u^\dagger )^k_i \\ \bar{d}_{Lk} ( V_d^\dagger )^k_i \end{pmatrix} \slashed{A}_\subh V^i_{Hj} Q_\subh^j ,
\end{equation}
where
\begin{equation}
	Q_\subh^i = \begin{pmatrix} u_\subh^i \\ d_\subh^i \end{pmatrix} .
\end{equation}
The rotations relevant to flavour physics are then
\begin{equation} \label{eq:fv1}
	( V_\subh^\dagger )^i_k \left( V_u \right)^k_j \equiv \left( V_{Hu} \right)^i_j, \qquad ( V_\subh^\dagger )^i_k \left( V_d \right)^k_j \equiv \left( V_{Hd} \right)^i_j ,
\end{equation} 
which are related through the Standard Model CKM matrix:
\begin{equation} \label{eq:fv2}
	V_{Hu}^\dagger V_{Hd} = V_\textrm{CKM} .
\end{equation}

For simplicity, in the following we will assume the matrix $k$ to be diagonal and flavour independent, forcing the T-odd fermions to be degenerate within different generations. The latter relations \eqref{eq:fv1} and \eqref{eq:fv2} then reduce to the usual SM definition of the CKM matrix, with the mirror fermion matrix $V_H$ as the identity matrix: this is called the minimal flavour violating scenario of LHT. For each generation of quarks and leptons, the following up- and down-type mass eigenstates are generated at $\mathcal{O} \left( v^{2}/f^{2} \right)$ via \eqref{eq:Todd}
\begin{equation}
	\boxed{m_{u_\subh} = \sqrt{2} k f \left( 1- \frac{1}{8} \frac{v^2}{f^2} \right)} \qquad \boxed{m_{d_\subh} = \sqrt{2} k f} .
\end{equation}
Hence, one obtains a total of twelve additional T-odd fermions, partners to the six quarks, the three charged leptons and the three neutrinos. One should notice that the up-type mass receives also a contribution from EWSB, since a coupling with the Higgs doublet is present in \eqref{eq:Todd} proportional to $c_{\xi}$ and $s_{\xi}$.

The next task is to write invariant Yukawa-like terms to give mass to the T-even (SM) combinations $u_{L+}$ and $d_{L+}$. In particular, in order to reduce the fine-tuning due to the SM top loop, the top Yukawa sector must realise a collective symmetry breaking pattern as well. One usually introduces the singlet fields $T_{L_1}$ and $T_{L_2}$ (and their right-handed counterparts) which are embedded, together with the previously defined $q_1$ and $q_2$ doublets, into the complete $SU(5)$ multiplets $\Psi_{1,t} = \left( q_1, T_{L_1}, \mathbf{0}_{1 \times 2} \right)^t$ and $\Psi_{2,t} = \left( \mathbf{0}_{1 \times 2}, T_{L_2}, q_2 \right)^t$. The $SU(5)$- and T-invariant Yukawa-like Lagrangian for the top sector then reads
\begin{align} \label{eq:tlht}
	\mathcal{L}_t = & - \frac{\lambda_1 f}{2 \sqrt{2}} \, \epsilon_{ijk} \, \epsilon_{xy} \left[ \left( \bar{\Psi}_{1,t} \right)_i \Sigma_{jx} \, \Sigma_{ky} - \left( \bar{\Psi}_{2,t} \, \langle \Sigma \rangle \right)_i \Sigma^\prime_{jx} \, \Sigma^\prime_{ky} \right] t^\prime_R \nonumber \\
	& - \lambda_2 f \left( \bar{T}_{L_1} T_{R_1} + \bar{T}_{L_2} T_{R_2} \right) \textrm{ + h.c.}
\end{align}
Here, the indices $i,j,k$ run over $1,2,3$ and $x,y$ over $4,5$, and $\Sigma^\prime = \langle \Sigma \rangle \, \Omega \, \Sigma^\dagger \, \Omega \, \langle \Sigma \rangle$ is the image of $\Sigma$ under T-parity \eqref{eq:sigmat}. Under T-parity, the new singlet fields transform as
\begin{align}
	\qquad & T_{L_1} \leftrightarrow - T_{L_2} \nonumber \\
	T: \qquad & T_{R_1} \leftrightarrow - T_{R_2} \nonumber \\
	\qquad & t^\prime_R \rightarrow t^\prime_R .
\end{align} 
The presence in \eqref{eq:tlht} of two different couplings $\lambda_1$ and $\lambda_2$ is due to the collective symmetry breaking mechanism.

The top Lagrangian \eqref{eq:tlht} finally contains the following terms
\begin{equation} 	\label{eq:toplht}
	\mathcal{L}_t \supset - \lambda_1 f \left( \frac{s_\Sigma}{\sqrt{2}} \, \bar{t}_{L+} \, t_R^\prime + \frac{1+c_\Sigma}{2} \, \bar{T}^\prime_{L+} \, t_R^\prime \right)- \lambda_2 f \left( \bar{T}^\prime_{L+} \, T^\prime_{R+} + \bar{T}^\prime_{L-} \, T^\prime_{R-} \right) \textrm{ + h.c.} \; ,
\end{equation}
for which $c_\Sigma = \cos \left( \sqrt{2} h / f \right)$, $s_\Sigma = \sin \left( \sqrt{2} h / f \right)$, and where we defined the T-parity eigenstates $t_{L+} = \left( t_{L_1} - t_{L_2} \right) / \sqrt{2}$ as before, and $T^\prime_{L\pm} = \left( T_{L_1} \mp T_{L_2} \right) / \sqrt{2}$, $T^\prime_{R\pm} = \left( T_{R_1} \mp T_{R_2} \right) / \sqrt{2}$. Among the terms that we have neglected in \eqref{eq:toplht}, there are the interaction terms of the T-odd eigenstate $t_{L-}$, which does not acquire any mass term from $\mathcal{L}_t$ while obtaining its mass from $\mathcal{L}_k$ as explained before. In $\mathcal{L}_t$ a different T-odd Dirac fermion $T_- \equiv \left( T^\prime_{L-},T^\prime_{R-} \right)$ obtains a high-scale mass 
\begin{equation} \label{eq:masst-}
	\boxed{m_{T_-} = \lambda_{2} \, f} .
\end{equation}  
The T-even combinations in $\mathcal{L}_t$, these are $\left( t_{L+}, t_R^\prime \right)$ and $\left( T_{L+}^\prime, T_{R+}^\prime \right)$, mix among each other: 
\begin{equation} 	\label{eq:matrix}
	-\mathcal{L}_t \supset \begin{pmatrix} \bar{t}_{L+} & \bar{T}_{L+}^\prime \end{pmatrix} \, \mathcal{M} \begin{pmatrix} t_R^\prime \\ T_{R+}^\prime \end{pmatrix} \textrm{ + h.c.} \quad \textrm{where} \quad \mathcal{M} = \begin{pmatrix}
		\frac{ \lambda_1 f}{\sqrt{2}} \sin \left( \frac{\sqrt{2} h}{f} \right) & 0 \\ \lambda_1 f \cos^2 \left( \frac{h}{\sqrt{2} f} \right) & \lambda_2 f \end{pmatrix} . 
\end{equation}
The mass terms are diagonalised by defining the linear combinations
\begin{align}
	t_L = & \cos \beta \cdot t_{L+} - \sin \beta \cdot T^\prime_{L+}
        & T_{L+} = \sin \beta \cdot t_{L+} + \cos \beta \cdot T^\prime_{L+} \nonumber \\ 
  t_R = & \cos \alpha \cdot t_R^\prime - \sin \alpha \cdot T^\prime_{L+}
        & T_{R+} = \sin \alpha \cdot t_R^\prime + \cos \alpha \cdot T_{R+}^\prime ,
\end{align}
where we used the dimensionless ratio $R = \lambda_1 / \lambda_2$ as well as the leading order expressions of the mixing angles 
\begin{equation}
	\sin \alpha = \frac{R}{\sqrt{1 + R^2}} \equiv \sqrt{x_L}, \qquad \sin \beta = \frac{R^2}{1+R^2} \frac{v}{f} \equiv x_L \frac{v}{f}. 
\end{equation}
Considering only the largest corrections induced by EWSB, the mass spectrum is given by
\begin{equation}
	m_t = \lambda_2 \, x_L \, v \left[ 1 + \frac{v^2}{f^2} \left( -\frac{1}{3} + \frac{1}{2} x_L \left( 1 - x_L \right) \right) \right] 
\end{equation}
and
\begin{equation} \label{eq:masst+}
	\boxed{m_{T_+} = \frac{f}{v} \frac{m_t}{\sqrt{x_L \left( 1-x_L \right)}} \left[ 1 + \frac{v^2}{f^2} \left( \frac{1}{3} - x_L \left( 1-x_L \right) \right) \right]} .
\end{equation}
$R$ and $\lambda_2$ are considered to be free parameters. However we can fix $\lambda_2$ requiring that, for given $\left( f, R \right)$, $m_t$ corresponds to the experimental top mass value: this way, the only free parameters in the T-even top sector are $f$ and $R$. 

One should mention that in reference \cite{Belyaev:2006jh} the authors have performed a study to fix the allowed range for $R$: by calculating the $J=1$ partial-wave amplitudes in the coupled system of ($t\bar{t}$, $T\bar{T}_+$, $b \bar{b}$, $WW$, $Z h$) states to estimate the tree level unitarity limit of the corresponding scattering amplitudes: the reported upper bound is 
\begin{equation} \label{eq:Rlim}
	\boxed{R \lesssim 3.3} .
\end{equation}

The other two generations of T-even up-type quarks acquire their mass through analogous terms as \eqref{eq:tlht}, but with the $T_{L_{1,2}}$ missing, since the Yukawa couplings are small and one does not have to worry about the quadratic divergences: no additional partners are then introduced in the spectrum besides the T-odd fermion $u_\subh$ which acquire mass via \eqref{eq:Todd}. 

Regarding the Yukawa interaction for the down-type quarks and charged leptons, two possible constructions of T-invariant Lagrangians are commonly known \cite{Chen:2006cs} and usually denoted as \emph{Case A} and \emph{Case B}, respectively. No additional partners are introduced as the quadratic corrections to the Higgs mass are negligible and do not require the introduction of the collective symmetry breaking mechanism in this sector. A prototype Lagrangian is given by \cite{Chen:2006cs},
\begin{equation}
	\label{eq:downlht}
	\mathcal{L}_d = - \frac{i \lambda_d f}{2 \sqrt{2}} \, \epsilon_{ij} \, \epsilon_{xyz} \left[ \left( \bar{\Psi}_{2}^\prime \right)_x \Sigma_{iy} \, \Sigma_{jz} \, X - \left( \bar{\Psi}_{1}^\prime \, \langle \Sigma \rangle \right)_x \Sigma^\prime_{iy} \, \Sigma^\prime_{jz} \, X^\prime \right] d^\prime_R \; ,
\end{equation}
with the same notation as in \eqref{eq:tlht}, and $\Psi_{1,2}^\prime$ the T-parity images of $\Psi_{1,2}$ \eqref{eq:quarklht}. $X$ is needed to achieve gauge invariance, transforming as a singlet under both $SU(2)_{1,2}$ and with $U(1)_{1,2}$ charges ($1/10,-1/10$), while $X^\prime$ is the image of $X$ under T-parity. Two choices are indeed possible for $X$, corresponding to the previously mentioned \emph{Case A} and \emph{Case B} respectively, namely $X=\left( \Sigma_{33} \right)^{-1/4} $ and $X = \left( \Sigma_{33} \right)^{1/4}$, where $\Sigma_{33}$ is the ($3,3$) component of the non-linear sigma model field $\Sigma$.

The free parameters of this sector are fixed in order to reproduce the SM masses. It turns out that the down-type and charged lepton couplings to the Higgs get corrections of order $\mathcal{O} \left( v^2/f^2 \right)$ with respect to their SM values, in the expansion of the non-linear sigma model. Furthermore, a higher suppression is registered in the $\emph{Case B}$ implementation, namely
\begin{align} \label{eq:caseb}
	\frac{g_{hd\bar{d}}}{g_{hd\bar{d}}^\mathrm{SM}} & = 1 - \frac{1}{4} \frac{v_\mathrm{SM}^2}{f^2} + \mathcal{O} \left( \frac{v_\mathrm{SM}^4}{f^4} \right) \quad \textrm{Case A} \\
	\frac{g_{hd\bar{d}}}{g_{hd\bar{d}}^\mathrm{SM}} & = 1 - \frac{5}{4} \frac{v_\mathrm{SM}^2}{f^2} + \mathcal{O} \left( \frac{v_\mathrm{SM}^4}{f^4} \right) \quad \textrm{Case B} ,
\end{align}
where we defined the Higgs coupling as $\mathcal{L}_d \supset g_{hd\bar{d}} \, hd\bar{d}$. The SM value is clearly $g_{hd\bar{d}}^\mathrm{SM} = m_d / v_\mathrm{SM}$.

Since the bottom coupling to the Higgs is highly relevant for the Higgs phenomenology, a different pattern is expected from the two different down-type Yukawa implementations, providing a distinctive phenomenology in the Higgs sector \cite{Chen:2006cs,Reuter:2012sd}. On the other hand, the Higgs phenomenology has a rather small impact on the topologies considered in direct searches. For this reason, we will focus only on the \emph{Case A} implementation throughout the paper. For sake of completeness, the results of the Higgs and EWPT combined analysis for \emph{Case B} will be provided in the appendix \ref{sec:higgsewptcaseb}.

\section{LHT phenomenology}
\label{sec:lhtphenomenology}
As detailed in section \ref{sec:lhtmodel}, under few assumptions involving mainly flavour independence in the mirror fermion sector, the model can be parametrised by only three free parameters. The parameter $f$ is the analogue of the pion decay constant in low-energy QCD and signifies the scale at which the global symmetry in the strong sector is spontaneously broken. Moreover $f$, or rather $(v/f)^2$, determines the amount of fine-tuning needed in the Higgs potential to stabilise against loop corrections. As Little Higgs theories were designed to overcome the little hierarchy problem, it is natural to demand a small fine-tuning and therefore a relatively low value of $f$. For example a scale $f \approx 2 \, \textrm{TeV}$ implies a fine-tuning of the order of $1 \%$. Of course, the very definition of fine-tuning has not an absolute physical meaning, and the interpretation of fine-tuning is also not totally physical. We leave it to the reader to judge whether a fine-tuning stronger than $1 \%$ would still be considered natural or not. Though it is clear that only with the full $14 \, \textrm{TeV}$ run one can enter contrived territories.

To become more specific, the naturalness of the model is usually quantified observing by how much the contributions from the heavy states ($\delta \mu^2$) exceed the observed value of the Higgs mass squared parameter ($\mu_{\text{obs}}^2$), as originally proposed in \cite{ArkaniHamed:2002qy}: 
\begin{equation} \label{eq:finetun}
	\Delta=\frac{|\delta \mu^2|}{\mu_{\text{obs}}^2}, \qquad \mu_{\text{obs}}^2= \frac{m_h^2}{2}.
\end{equation}
For example, if the new contributions to the Higgs mass squared parameter exceed $\mu_{\text{obs}}^2$ by a factor of 5, i.e. $\Delta =5$, one says that the model requires $20\%$ of fine-tuning. Clearly, the lower the value of fine-tuning, the worse is the naturalness of the model. The dominant log-divergent contribution to the Higgs mass squared parameter comes from the top and its $T_{+}$ heavy partner loops, and is given by \cite{ArkaniHamed:2002qy} 
\begin{equation}
	\delta \mu^2 = -\frac{3 \lambda_t^2 m_T^2}{8 \pi^2} \log{\frac{\Lambda^2}{m_T^2}}
\end{equation}
where $\Lambda=4 \pi f$ is the cut-off of the nonlinear sigma model, $\lambda_t$ is the SM top Yukawa coupling and $m_T$ is the mass of the heavy top partner. In the next sections we will thus quantify the required amount of fine-tuning using equation \eqref{eq:finetun}.

The other two parameters $k$ and $R = \lambda_1 / \lambda_2$ parametrise the couplings in the mirror fermion sector \eqref{eq:Todd} and in the top partner sector \eqref{eq:tlht}, respectively. Therefore to constrain the symmetry breaking scale $f$ in a consistent way, it is needed to exclude regions in parameter space while varying $k$ and $R$ within their theoretical or experimental bounds, see equations \eqref{eq:Rlim} and \eqref{eq:kfbound}. The model phenomenology could change drastically for different values of the latter two free parameters, as we will detail in the next sections.

\subsection{Particle spectrum and decay modes}
\label{sec:particlespectrum}
Generally speaking, the model is realised in such a way that only the new partners of the SM fields acquire a mass from the global (local) spontaneous symmetry breaking $SU(5) \rightarrow SO(5)$ ($[SU(2) \otimes U(1)]^{2} \rightarrow [SU(2) \otimes U(1)]_\textrm{diag}$) at the scale $f$, while the SM states remain massless. EWSB further generates corrections of order $v^2/f^2$ to the partner masses, and weak scale $v$ masses for the other SM states analogously to the original Higgs mechanism. 

Once the values of the gauge coupling constants and of the parameters in the scalar potential are fixed, the mass of the gauge boson- and scalar partners are completely determined by the scale $f$. General features are the identification of the heavy photon $A_\subh$ as the lightest T-odd particle, therefore being stable unless the quark partners $q_\subh$ become even lighter. This happens if
\begin{align}
	m_{u_\subh} < m_{A_\subh} & \quad \textrm{if} \quad k < \frac{g'}{\sqrt{10}} \left(1 - \frac{1}{2} \frac{v^2}{f^2} + \cdots \right) \nonumber \\ 
	m_{d_\subh} < m_{A_\subh} & \quad \textrm{if} \quad k < \frac{g'}{\sqrt{10}} \left(1 - \frac{5}{8} \frac{v^2}{f^2} + \cdots \right) 
\end{align}
which corresponds to values of $k \lesssim 0.1$. The heavy W and Z partners are degenerate up to corrections of order $v^2/f^2$, both being lighter than the different components of the complex triplet $\phi$, which are also degenerate at the order we are considering.

\begin{figure}[!ht]
	\begin{center}
		\includegraphics[scale=0.6]{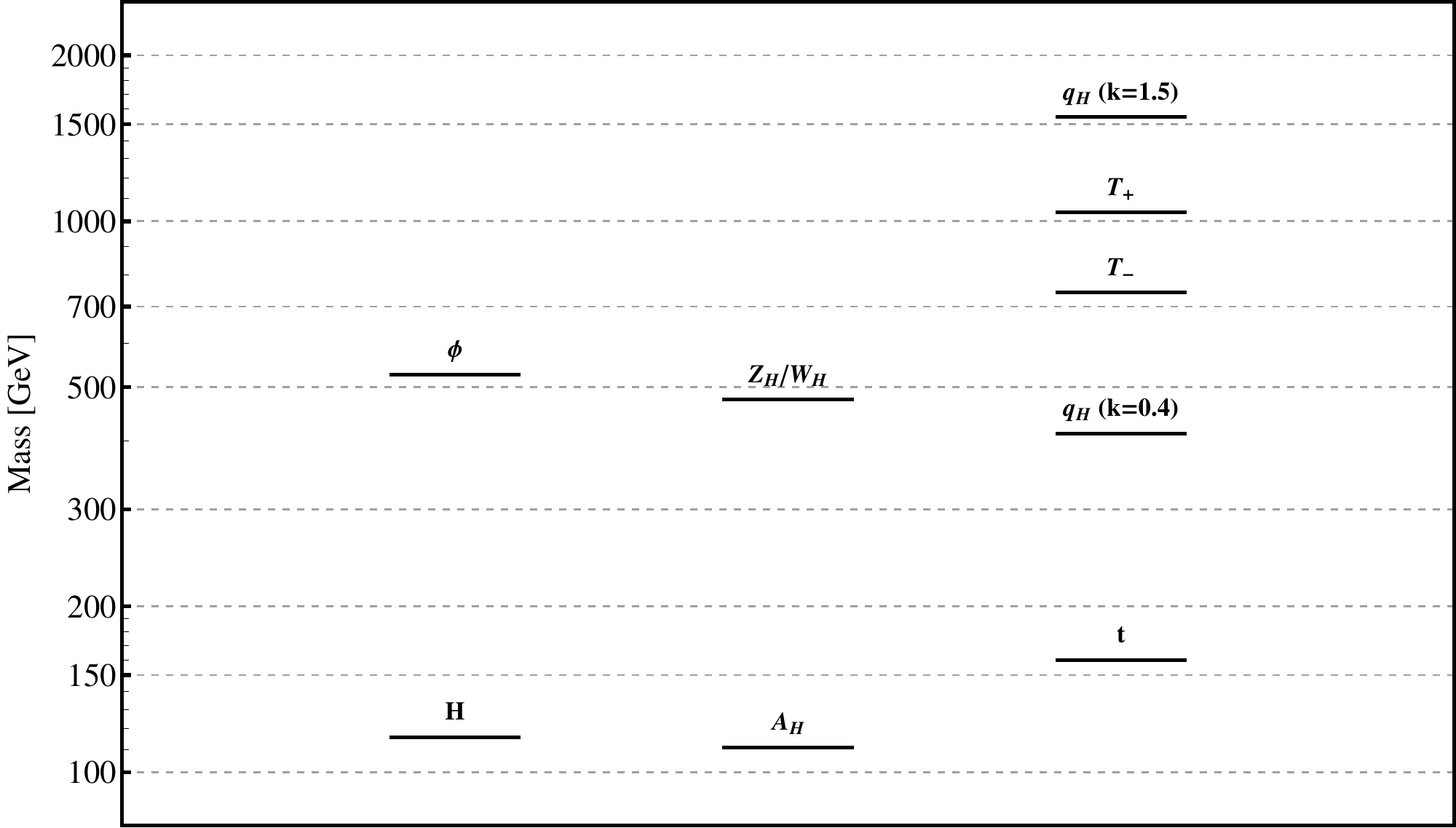} 
		\caption{LHT partner masses showing the effect of $k$ on the heavy quark masses. $f$ and $R$ are fixed to $800 \, \textrm{GeV}$ and $1.0$ respectively.}
		\label{fig:massspectrum}
	\end{center}
\end{figure}

On the other hand, the LHT model building requires the presence of other free parameters in the fermion sectors, namely $k$ and $R = \lambda_1 / \lambda_2$ as described before, making the fermion spectrum  dependent on those values besides the scale $f$. In particular the top partners $T_+, T_-$ are always heavier than all bosonic partners, as one can partially see from figure \ref{fig:massspectrum}, where we have fixed $R=1.0$ minimising the $T_+$ mass (\ref{eq:masst+}). The $T_-$ mass (\ref{eq:masst-}) is proportional to $R^{-1}$, but it is always heavier than all bosonic partners in the allowed range $R \lesssim 3.3$, too. Notice further that the T-even top partner $T_+$ is always heavier than its T-odd partner $T_-$, with their mass splitting proportional to $R$. The mass of the quark partners depends on the value of $k$, and they are heavier than all the gauge boson partners if
\begin{align}
	m_{u_\subh} > m_{W_\subh}, m_{Z_\subh} & \quad \textrm{if} \quad k > \frac{g}{\sqrt{2}} \nonumber \\ 
	m_{d_\subh} > m_{W_\subh}, m_{Z_\subh} & \quad \textrm{if} \quad k > \frac{g}{\sqrt{2}} \left(1 - \frac{1}{8} \frac{v^2}{f^2} + \cdots \right) .
\end{align}
This corresponds to $k \gtrsim 0.45$, making the decay $q_\subh \rightarrow V_\subh \; q$ kinematically allowed, where we defined $V_\subh \equiv W_\subh^\pm, Z_\subh$. For $k \lesssim 0.45$ the only kinematically allowed decay of the quark partners is $q_\subh \rightarrow A_\subh \; q$. A compressed mass spectrum is generated in the region $0.1 \lesssim k \lesssim 0.2$ where the mass difference between $q_\subh$ and $A_\subh$ is rather small. For even smaller values of $k$, namely $k \lesssim 0.1$, the quark partners become lighter than the heavy photon $A_\subh$ and therefore stable. We can safely say that this region could be considered as excluded or in high tension with the experimental observations. In particular R-hadron constraints from the LHC \cite{Aad:2012pra} and coloured particle constraints from cosmological observations \cite{Barbier:2004ez}, strongly disfavour stable charged particles. In figure \ref{fig:massspectrum} we plot typical mass spectra of the LHT partners for a reference values $f = 800 \, \textrm{GeV}$, $R=1.0$ and $k=1.5$ or $k=0.4$. 

\begin{table}[!ht]
	\centering
	\begin{tabular}{l l r r}
		\toprule[1pt]
		Particle & Decay & $\textrm{BR}_{k=1.0}$ & $\textrm{BR}_{k=0.4}$ \\
		\midrule[1pt]
		$d_\subh$ & $W_\subh^- \; u$ & $63\%$ & $0\%$ \\
		& $Z_\subh \; d$ & $31\%$ & $0\%$ \\
		& $A_\subh \; d$ & $6\%$ & $100\%$ \\
		\midrule 
		$u_\subh$ & $W_\subh^+ \; d$ & $61\%$ & $0\%$ \\
		& $Z_\subh \; u$ & $30\%$ & $0\%$ \\
		& $A_\subh \; u$ & $9\%$ & $100\%$ \\
		\midrule
		$T_\subh^+$ & $W^+ \; b$ & $46\%$ & $46\%$ \\
		& $Z \; t$ & $22\%$ & $22\%$ \\
		& $H \; t$ & $21\%$ & $21\%$ \\
		& $T_\subh^- \; A_\subh$ & $11\%$ & $11\%$ \\
		\midrule
		$T_\subh^-$ & $A_\subh \; t$ & $100\%$ & $100\%$ \\
		\midrule
		$\Phi^0$ & $A_\subh \; Z$ & $100\%$ & $100\%$ \\
		\midrule
		$\Phi^P$ & $A_\subh \; H$ & $100\%$ & $100\%$ \\
		\bottomrule[1pt]
	\end{tabular}
	\hspace{1mm}
	\begin{tabular}{l l r r}
		\toprule[1pt]
		Particle & Decay & $\textrm{BR}_{k=1.0}$ & $\textrm{BR}_{k=0.4}$ \\
		\midrule[1pt]
		$\Phi^\pm$ & $A_\subh \; W^\pm$ & $100\%$ & $100\%$ \\
		\midrule \vspace{0.5mm}
		$\Phi^{\pm\pm}$ & $A_\subh \; (W^\pm)^2$ & $99\%$ & $96\%$ \\
		\midrule \vspace{0.5mm}
		$A_\subh$ & stable & & \\
		\midrule \vspace{0.5mm}
		$W_\subh^\pm$ & $A_\subh \; W^\pm$ & $100\%$ & $2\%$ \\
		& $u_\subh \; d$ & $0\%$ & $44\%$ \\
		& $d_\subh \; u$ & $0\%$ & $27\%$ \\
		& $l_\subh^\pm \; \nu$ & $0\%$ & $13.5\%$ \\
		& $\nu_\subh \; l^\pm$ & $0\%$ & $13.5\%$ \\
		\midrule \vspace{0.5mm}
		$Z_\subh$ & $A_\subh \; H$ & $100\%$ & $2\%$ \\
		& $d_\subh \; d$ & $0\%$ & $40\%$ \\
		& $u_\subh \; u$ & $0\%$ & $30\%$ \\
		& $l_\subh^\pm \; l^\mp$ & $0\%$ & $14\%$ \\
		& $\nu_\subh \; \nu$ & $0\%$ & $14\%$ \\
		\bottomrule[1pt]
	\end{tabular}
	\caption{An overview of the decay modes with the corresponding branching rations of all new LHT particles for $f = 1 \, \textrm{TeV}$ and $R = 1.0$. Two scenarios are listed, where the heavy quarks $q_{\subh}$ are either lighter ($k = 0.4$) or heavier ($k=1.0$) than the boson partners. The heavy leptons decay analogously to the heavy quarks and the decays involving generic up or down quarks have to be considered as summed over all flavours.}
	\label{tab:lhtdecays}
\end{table}

Given the previous discussion, it is clear that the decay modes of the quark partners and of the gauge boson partners mostly depend on the value of $k$. All branching ratios have indeed a mild dependence on $f$ and $R$. In table \ref{tab:lhtdecays} we present the typical branching ratios for two different benchmark scenarios, namely $k=1.0$ and $k=0.4$ with $f = 1 \, \textrm{TeV}$, $R=1.0$.

\subsection{Production modes and experimental signatures}
\label{sec:productionmodes}

In this section we will discuss, in order of decreasing cross section at the LHC, the production of the different LHT new particles, updating the results presented in \cite{Belyaev:2006jh}. Notice that due to T-parity, only the T-even top partner $T_+$ could be singly produced, while all other particles have to be pair produced, highly reducing the available phase space with increasing masses. The plots in this section depicting the production cross sections as a function of the symmetry breaking scale $f$ are done for the benchmark point $R = 1.0$ and $k = 1.0$, unless stated otherwise. 

\begin{figure}[!ht]
	\begin{center}
		\includegraphics[scale=0.7]{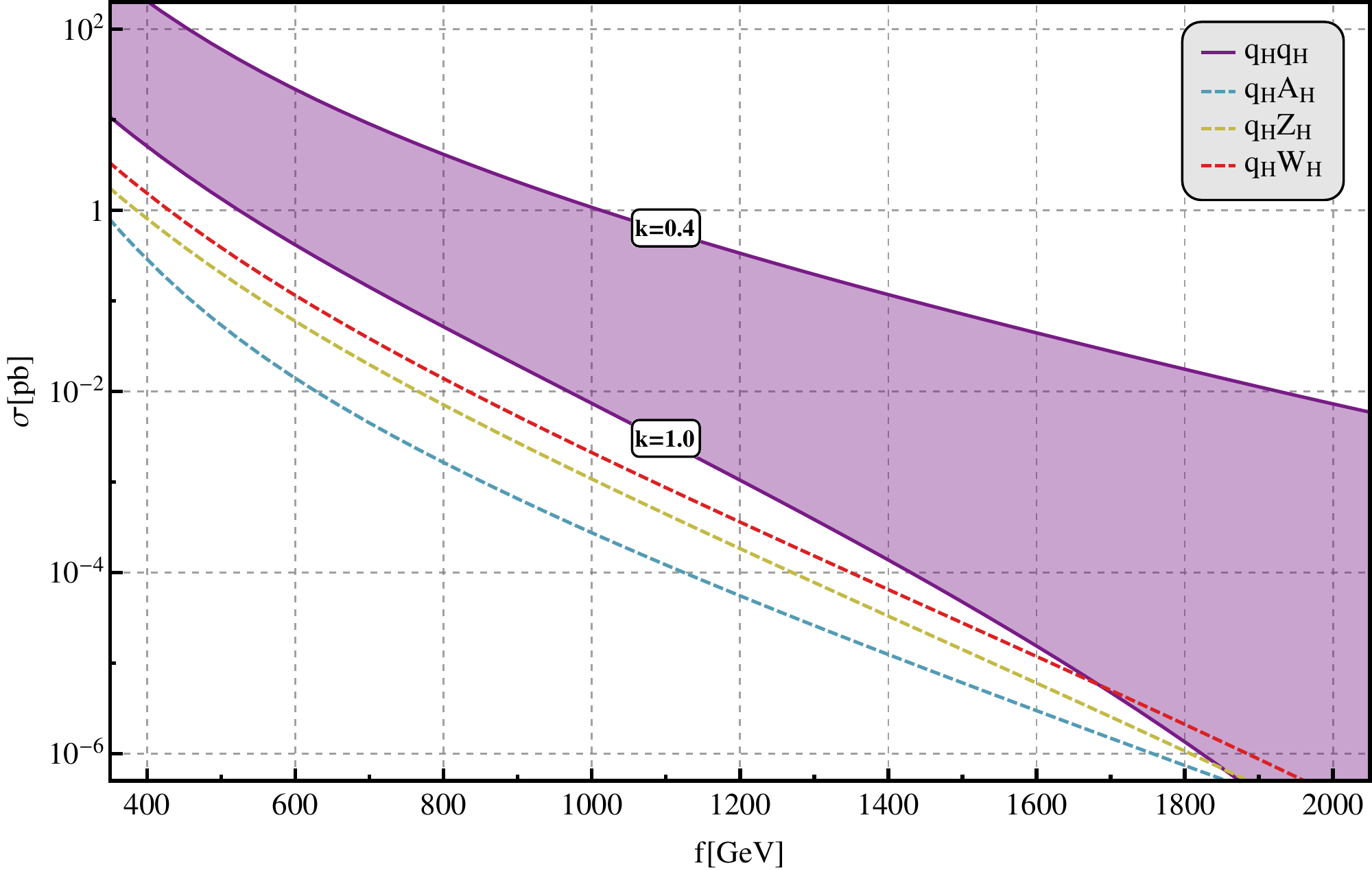} 
		\caption{Pair- and associated production cross section of the quark partners $q_\subh$ at the LHC operating at $8 \, \textrm{TeV}$, for reference values of $R=1.0$ and $k=1.0$. The pair production line width corresponds to values of $k \in [0.4,1.0]$.}
		\label{fig:heavyquarkproduction}
	\end{center}
\end{figure}

Since the LHC is a hadron collider, the pair production of quark partners $q_\subh$ will be significant, especially if their masses are not too large. Opposite sign quark partners can be dominantly produced via QCD processes, but also via electroweak processes involving a heavy $W_\subh$ or $Z_\subh$, $A_\subh$ in the t-channel. Among the production of a quark partner in association with a gauge boson partner, the dominant contribution comes from the associated production with a heavy $W_\subh$, because of the different strength of the couplings between $q_\subh$ and $V_\subh$. In figure \ref{fig:heavyquarkproduction} we plot pair and associated productions of quark partners at LHC8. Since the mass of the quark partners is proportional to $k$, the $q_\subh$ pair production is expected to decrease faster with respect to the associated productions for higher values of $k$: the width of the pair production line corresponds indeed to values of $k \in [0.4,1.0]$.

\begin{figure}[!ht]
	\begin{center}
		\includegraphics[scale=0.7]{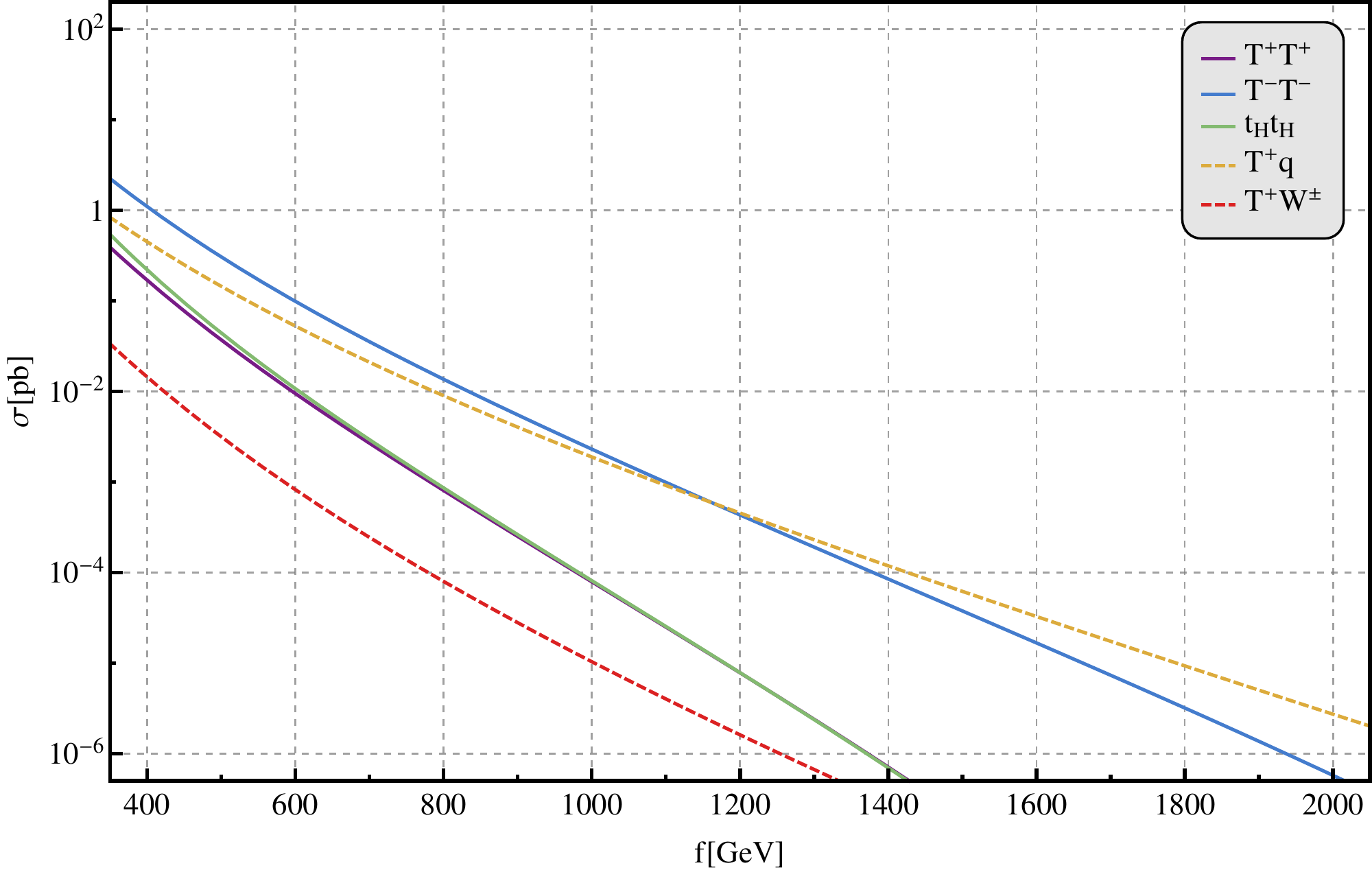} 
		\caption{Production cross section of LHT top partners at the LHC operating at $8$ TeV, for reference values of $R=1.0$ and $k=1.0$.}
		\label{fig:toppartnerproduction}
	\end{center}
\end{figure}

Because of the mass spectrum described in section \ref{sec:particlespectrum}, the T-odd $T_-$ has the largest pair production cross section compared with the pair production of $T_+$ and $t_\subh$ (for $k \gtrsim 0.7$). Clearly, lower values of $k$ reduce the mass of the quark partners $q_\subh$, making their pair production the dominant process. With increasing values of $f$, both $T_+$ and $T_-$ become heavier, making the single production of the $T_+$ in association with a light quark (through a diagram involving a t-channel $W$ with an initial state bottom quark) comparable in size or even larger than the $T_-$ pair production. The dominant associated production of the $T_+$ with a SM gauge boson is the one involving the $W^\pm$, which is suppressed with respect to the other production modes since the $b T_+ W$ coupling is proportional to $v/f$. 

The qualitative behaviour described above can be slightly different by changing the values of $k$ and $R$. In particular $R \ll 1$ can be considered as the decoupling limit of both $T_+$ and $T_-$, making both pair and associated productions vanishing, while for $R > 1$ the mass splitting between $T_-$ and $T_+$ increases, making the $T_-$ pair production sizeably larger than the associated $T_+ \; q$ production. In figure \ref{fig:toppartnerproduction} we plot the different production cross sections of the top partners $T_+$ and $T_-$ fixing $R=1.0$ and $k=1.0$.

\begin{figure}[!ht]
	\begin{center}
		\includegraphics[scale=0.7]{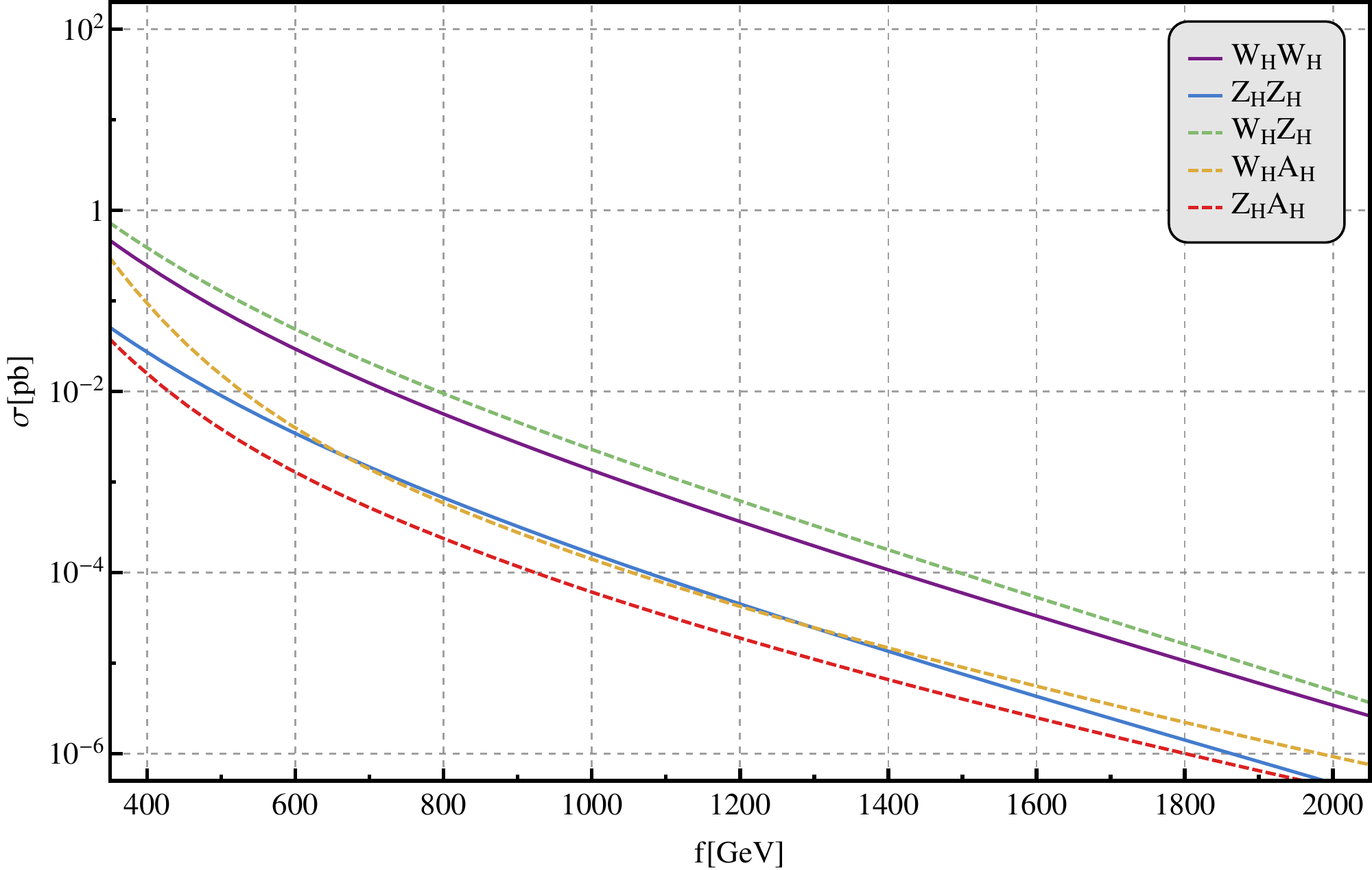}
		\caption{Production cross section of LHT gauge bosons partners at the LHC operating at $8 \, \textrm{TeV}$, for reference values of $R=1.0$ and $k=1.0$.}
		\label{fig:heavygaugebosonproduction}
	\end{center}
\end{figure}

With generally smaller cross section, the different production modes for pairs of heavy gauge bosons $V_\subh \; V_\subh$, with $V_\subh \equiv W_\subh^\pm, Z_\subh, A_\subh$ are plotted in figure \ref{fig:heavygaugebosonproduction}. Their dependence on the parameters is smoother with respect to the fermion production modes, affecting only the masses of t-channel exchanged fermionic partners. The $V_\subh \; V_\subh$ pair production is indeed generated via s-channel exchange of SM gauge bosons or via t-channel exchange of fermionic partners.

The production modes involving the heavy triplet scalar components ($\phi^0$, $\phi^P$, $\phi^+$, $\phi^{++}$) will not be considered here, neither the ones involving the lepton partners $l_\subh$, $\nu_\subh$. Due to the fact that these production cross sections are parametrically smaller, they are therefore not affecting the LHC phenomenology relevant for our studies. Given the production and decay modes, it is straightforward to categorise the relevant signatures of the model with respect to the LHC searches. We present a table in appendix \ref{sec:lhttopologies}.

\section{Experimental searches}
\label{sec:experimentalsearches}
In this section, we first discuss the electroweak precision constraints, then the Higgs data, and finally the direct LHC searches for new particles.

\subsection{Electroweak precision observables}
\label{sec:electroweakprecisionobservables}
Historically, the most severe constraints on the parameter space of the different implementations of the Little Higgs paradigm have always arisen from EWPT \cite{Csaki:2002qg,Kilian:2003xt,Hewett:2002px}. The most serious constraints resulted from tree level corrections to precision electroweak observables due to exchanges of additional heavy gauge bosons present in the theories. In the class of product group models, this has been indeed the reason for introducing T-parity which exchanges the gauge groups, as explained in section \ref{sec:gaugesector}, making almost all new particles T-odd and all SM particles T-even. Tree level couplings of light states with only one heavy particle are thus forbidden, and no large contributions from higher dimensional operators obtained by integrating out the heavy fields are generated. The lower bound on the symmetry breaking scale $f$ from EWPT is then relaxed, making the new particles eventually observable at the LHC. On the other hand, one has to pay the price for pair production.

In the LHT model in particular, the only new particle which is T-even is the T-even top partner $T_+$. However it can contribute at tree level only to observables involving the SM top quark, such as its couplings to $W$ and $Z$ bosons: since these couplings have not been measured experimentally yet, no constraints arise at tree level from the T-even top partner as well. The leading contributions to electroweak observables arise therefore from one loop diagrams involving the new T-even and T-odd states. We refer to the literature \cite{Hubisz:2005tx,Asano:2006nr,Berger:2012ec,Reuter:2012sd} for a comprehensive review.

\begin{figure}[!ht]
	\begin{center}
		\includegraphics[scale=0.6]{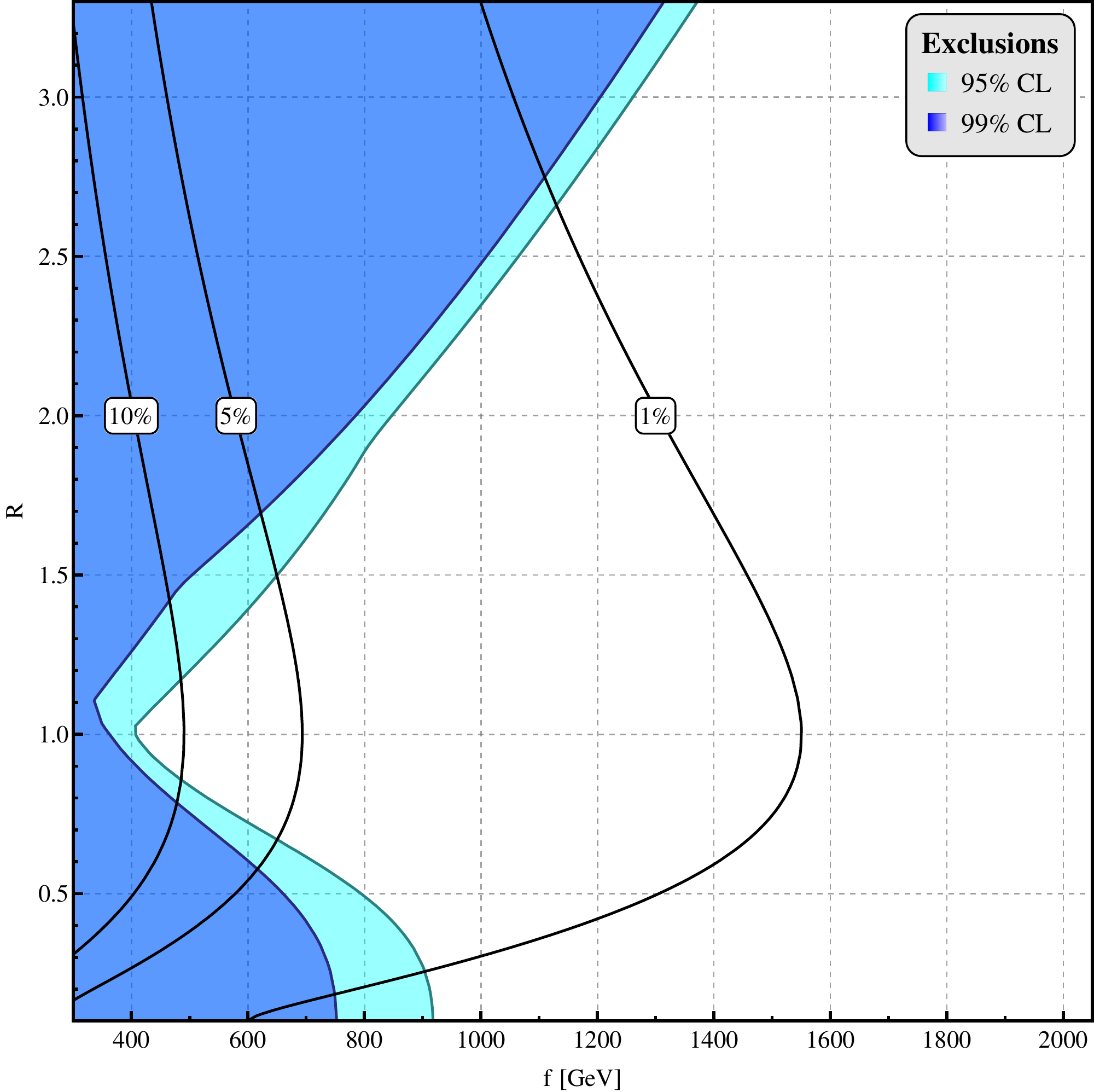} 
		\caption{Excluded parameter space regions at $95\%$ and $99\%$ CL from EWPT. The thick black lines represent contours of required fine-tuning.}
		\label{fig:EWPT}
	\end{center}
\end{figure}

Following most of the details of the analysis realised in \cite{Reuter:2012sd}, including 21 different low-energy and $Z$-pole precision observables for $m_h = 124.5 \, \textrm{GeV}$ \cite{Beringer:1900zz}, a $\chi^2$ analysis in the ($f,R$) plane results in a lower bound on the symmetry breaking scale 
\begin{equation} \label{eq:EWPTlimit}
	\boxed{f \gtrsim \limitEPWT \, \textrm{GeV}} \quad \textrm{at} \quad 95\% \textrm{ CL} ,
\end{equation}
compare with figure \ref{fig:EWPT}. In this updated analysis we included also the T-odd fermion contributions to the $T$ parameter and fit the value of $k$ minimising the $\chi^2$ for each point in the ($f,R$) parameter space, letting $k$ range between the lower bound arising from the direct searches presented in section \ref{sec:directsearches} --- which is at least $k=0.6$ for any given $f$ --- and the effective operator bound to be discussed below in equation \eqref{eq:kfbound}. Note that the rather low value in equation \eqref{eq:EWPTlimit} results from the dip around $R \sim 1$, where the LHT contributions to the EWPT are minimised. The value in equation \eqref{eq:EWPTlimit} is the overall exclusion limit at $95\%$ CL, independent of $R$.

The thick black lines of figure \ref{fig:EWPT} enclose regions of required fine-tuning (on the left hand side of the lines) as defined at the beginning of section \ref{sec:lhtphenomenology}, and the level of fine-tuning is also denoted in the plot. 

\subsection{Higgs searches}
\label{sec:higgssearches}
Since the discovery of a bosonic resonance, we have entered a new era of Higgs physics. Besides EWPT, flavour constraints and direct searches of particles, the Higgs sector has become a useful framework for testing the validity of BSM models. 

It is customary for the experimental collaborations to express the results of the SM-like Higgs searches in terms of a signal strength modifier $\mu$, defined as the factor by which the SM Higgs signal is modified for a given value of $m_h$: 
\begin{equation} \label{eq:mu}
	\mu^i = \frac{n_S^i}{n^{\textrm{SM}, \, i}_S} = \frac{\sum_p \sigma_p \cdot \epsilon_i^p}{\sum_p \sigma_p^\textrm{SM} \cdot \epsilon_i^p} \cdot \frac{\textrm{BR}_i}{\textrm{BR}_i^\textrm{SM}} 
\end{equation}
where $i$, $p$ refer to a specific Higgs decay channel and production mode, respectively. Furthermore $n_S^i$ is the total number of expected Higgs signal events evaluated in a chosen model passing the selection cuts, and $\epsilon_i^p$ is the cut efficiency for a given process ($p,i$). For each Higgs decay channel considered, the ATLAS and CMS collaborations report the best-fit value $\hat{\mu}$ for a given hypothesis on $m_h$, while the cut efficiencies (or equivalently the signal composition in terms of the different production modes) are instead only partially reported.

Considering a generic Higgs process ($p,i$) the cut efficiency $\epsilon_i^p$ can indeed be expressed as
\begin{equation} \label{eq:efficiency}
	\epsilon_i^p = \frac{n^{\textrm{SM}, \; i}_S \cdot \zeta_i^p}{\mathcal{L} \cdot \sigma_p^\textrm{SM} \cdot \textrm{BR}_i^\textrm{SM}}
\end{equation}
where $n^{\textrm{SM}, \, i}_S \cdot \zeta_i^p$ is the fraction of the SM expected signal events produced via the process ($p,i$) passing the selection cuts, $\mathcal{L}$ is the integrated luminosity, and $\sigma_p^\textrm{SM}$, $\textrm{BR}_i^\textrm{SM}$ are the SM cross section and branching ratio of the considered process ($p,i$), respectively. One should notice that $\sum_p \zeta_i^p = 1$, while $\sum_p \epsilon_i^p < 1$ in general. If the signal composition in terms of the different production modes ($\zeta_p^i, \; \forall \; p$) is reported, equation \eqref{eq:mu} then simplifies to
\begin{equation} \label{eq:musimplified}
	\mu^i = \left( c_g \cdot \zeta_g^i + c_V^2 \cdot \zeta_V^i + c_t^2 \cdot \zeta_t^i \right)  \frac{\textrm{BR}_i}{\textrm{BR}_i^\textrm{SM}}
\end{equation}
where $g,V,t$ refer to gluon, vector and top initiated productions respectively, and where
\begin{align}
	\sigma_g^\textrm{BSM} & = c_g \cdot \sigma_g^\textrm{SM}, \nonumber \\
	\sigma_V^\textrm{BSM} & = c_V^2 \cdot \sigma_V^\textrm{SM}, \nonumber \\
	\sigma_t^\textrm{BSM} & = c_t^2 \cdot \sigma_t^\textrm{SM}. 
\end{align}
The rescaling factors $c_g$, $c_V$ and $c_t$ are model dependent and parametrise the rescaling of the $h \rightarrow gg$ partial width and of the $hVV$ ($V \equiv W, Z$) or $h t \bar{t}$ vertices with respect to their SM values respectively, see \cite{Reuter:2012sd,Azatov:2012qz} for more details. For the channels where the signal composition is not reported, one is forced to neglect the efficiencies from equation \eqref{eq:mu}, thus making the BSM predictions less reliable.

\begin{figure}[!ht]
	\begin{center}
		\includegraphics[scale=0.6]{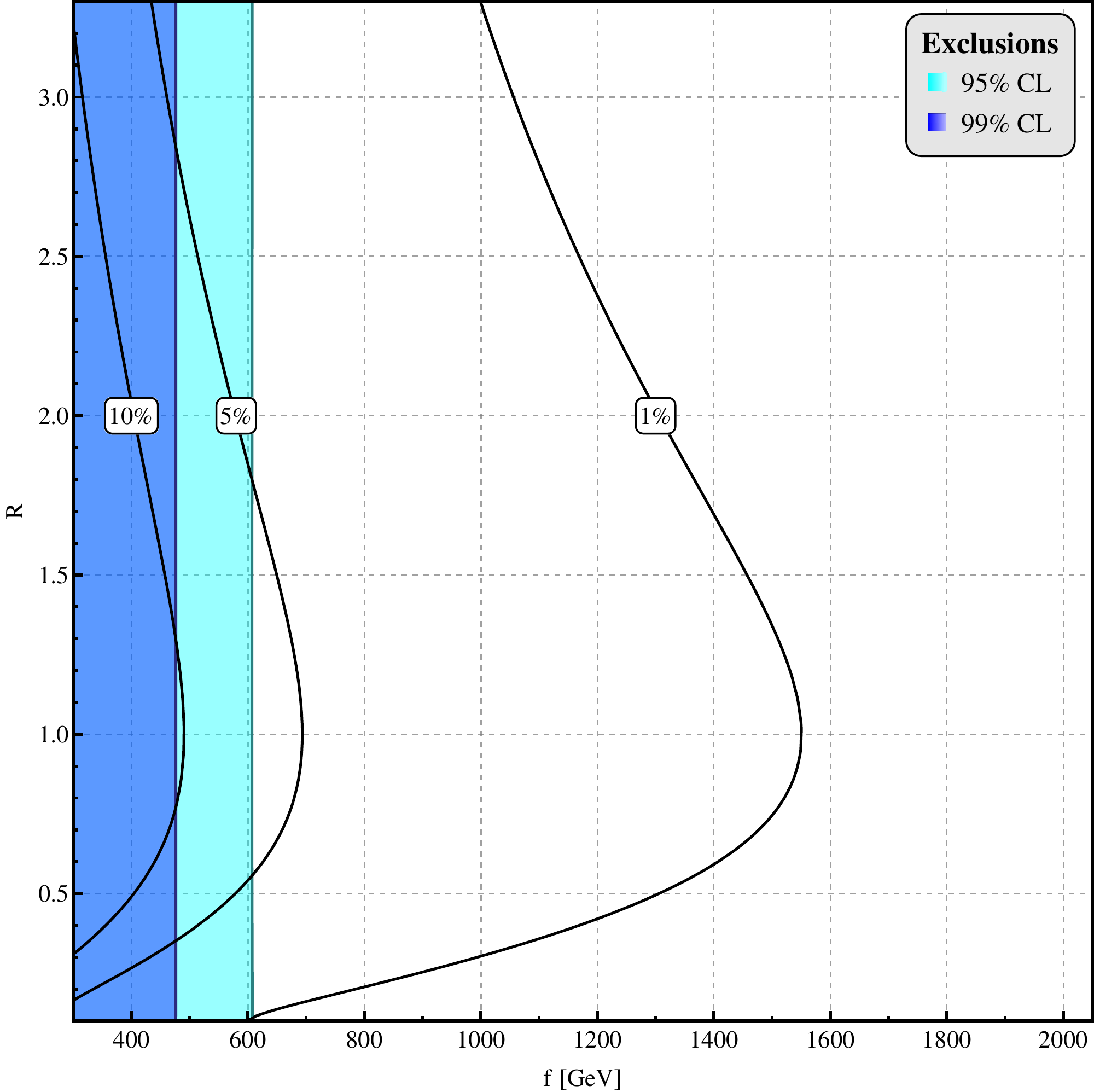} 
		\caption{Excluded parameter space regions at $95\%$ and $99\%$ CL from ATLAS and CMS $25 \, \textrm{fb}^{-1}$ Higgs searches. The thick black lines represent contours of required fine-tuning.}
		\label{fig:Higgs}
	\end{center}
\end{figure}

The most recent data made public by both collaborations cover up to $25 \, \textrm{fb}^{-1}$ analysed data for most of the $7 \! + \! 8 \, \textrm{TeV}$ Higgs searches. In this paper we report an update of the analysis in the ($f,R$) plane realised in \cite{Reuter:2012sd} with the updated dataset as first summarised in \cite{Azatov:2012qz}. For completeness we report the explicit values used in the fit in appendix \ref{sec:higgsdata}. Unlike in the original analysis \cite{Reuter:2012sd}, we do not reconstruct the $8 \, \textrm{TeV}$ likelihood functions when only the $7 \, \textrm{TeV}$ and combined $7 \! + \! 8 \, \textrm{TeV}$ results are reported, while we use the $7 \! + \! 8 \, \textrm{TeV}$ data as if it were all coming from an $8 \, \textrm{TeV}$ run, as suggested in \cite{Azatov:2012qz}. The only error incurred doing this is from a different weighting that would arise in the separate production modes, but this fractional difference is negligible.

The updated lower bound on the symmetry breaking scale, obtained from figure \ref{fig:Higgs}, is
\begin{equation}
	\boxed{f \gtrsim \limitHiggs \, \textrm{GeV}} \quad \textrm{at} \quad 95\% \textrm{ CL} .
\end{equation}
One should notice that with the inclusion of the $25 \, \textrm{fb}^{-1}$ dataset, Higgs searches have finally overwhelmed EWPT in driving the lower bound on the LHT symmetry breaking scale $f$, at least in the region around $R \sim 1$, where the EWPT exclusion is least. The regions of required fine-tuning are also presented in the plot.

\subsection{Direct LHC searches}
\label{sec:directsearches}
In this section we discuss the impact of direct LHC searches from the $8 \, \textrm{TeV}$ run for the Littlest Higgs with T-parity. To obtain the exclusion limits from the recasting analysis we first implemented the LHT model in FeynRules \cite{Christensen:2008py} combining the Feynman rules presented in \cite{Han:2003wu,Hubisz:2004ft,Blanke:2006eb}.\footnote{The FeynRules model implementation is available upon request by the authors.} The FeynRules package has been used to export the model to the UFO \cite{Degrande:2011ua} format in order to interface it with the MadGraph \cite{Alwall:2011uj} Monte Carlo generator. The model is then validated by reproducing the known results from the literature for both production cross sections and decay of the heavy particles in the model. Furthermore, cross checks of the implementation with the event generator WHIZARD \cite{Kilian:2007gr,Moretti:2001zz} and its FeynRules interface \cite{Christensen:2010wz} have been done. The results in section \ref{sec:lhtphenomenology} are in agreement with the established literature \cite{Belyaev:2006jh,Hubisz:2004ft} on LHT models.

MadGraph is used to generate parton level events which are then interfaced into the Pythia 6.42 \cite{Sjostrand:2006za} parton shower. The result is further processed in Delphes 3.0 \cite{deFavereau:2013fsa} to simulate either the ATLAS or CMS detector in a fast manner. Different analyses published by ATLAS and CMS can then be recasted for the LHT model to extract exclusion limits. In particular, for each considered analysis, we evaluated the efficiencies of the analysis-dependent cuts applied to a LHT signal which could mimic the experimental final state topology under consideration.\footnote{A Mathematica package for this purpose has been developed and is also available upon request.} The predicted visible cross section is then simply given by a reweighting of the signal cross section times the evaluated efficiency. The experimental $95\%$ CL upper bound on the visible cross section can finally be used to determine the possible exclusion of the corresponding parameter space point.

Since most of the final states mimic supersymmetry final states with significant amounts of missing transverse energy, we mostly discuss these searches in the following paragraphs. However, we begin with a paragraph on constraints from effective operators bounds. A phenomenologically interesting feature of the LHT model is the power counting of $k$ which leads to an upper bound and not a lower bound for the particles running in the loop.  Then we discuss the supersymmetry searches by ATLAS and CMS bearing in mind the determination of the lower exclusion limit on the scale $f$. This is most easily done in processes where only the parameters $f$ and $k$ play a role and the exclusion limits can be given in the $(f,k)$ plane. These are then summarised in the next section.

\paragraph{Effective operator bounds:}
The T-odd quark partners of the SM fermions can generate four-fermion operators via box diagrams involving the exchange of NGBs \cite{Hubisz:2005tx}. Assuming a diagonal and flavour-independent matrix $k$, the following set of four fermion operators is generated 
\begin{equation}
	\mathcal{O}_\textrm{4-f} = - \frac{k^2}{128 \, \pi^2 f^2} \bar{\psi}_L \gamma^{\mu} \psi_L \bar{\psi}^\prime_L \gamma_{\mu} \psi^\prime_L + \mathcal{O} \left( \frac{g}{k} \right) , 
\end{equation}
where $\psi$ and $\psi^{\prime}$ are (possibly distinct) SM fermions. On the other hand these four fermion operators may also be generated through strongly coupled physics above the scale $\Lambda = 4 \pi f$. An estimate for these contributions is
\begin{equation}
	\mathcal{O}_{\Lambda} \approx \pm \frac{C_\Lambda}{16 \, \pi^2 \Lambda^2} \bar{\psi}_L \gamma^\mu \psi_L \bar{\psi}^\prime_L \gamma_\mu \psi^\prime_L , 
\end{equation}
where the coupling $C_\Lambda$ should be roughly $\mathcal{O}(1)$.

Experimental bounds on four fermion interactions provide an upper bound on the T-odd fermion masses, which then yield an upper bound on $k$ for a given value of $f$. Possible constraints at the LHC come from operators involving four quarks, for example searches in the angular distribution of dijets \cite{ATLAS:2012pu,Chatrchyan:2012bf}. These experimental searches give constraints on the operator coefficient in the range of $\Lambda = 15 \, \textrm{TeV}$ for constructive interference which we are considering here. Although these searches are promising candidates to further constrain the parameter $k$, the most stringent bounds are actually still from LEP searches. The strongest constraint comes from the \emph{eedd} operator $\Lambda_\textrm{4-f} = 26.4 \, \textrm{TeV}$ \cite{Hubisz:2005tx,Beringer:1900zz}. This requires the coefficient of the four fermion operator to be smaller than $2 \pi /\Lambda_\textrm{4-f}^2$. This yields the following upper bound for $k$ 
\begin{equation} 	\label{eq:kfbound}
	k^2 < 256 \, \pi^3 \frac{f^2}{\Lambda_\textrm{4-f}^2} \pm \frac{C_\Lambda}{2 \pi^2} . 
\end{equation}
This bound is plotted in the total exclusion plot at the end of this section in figure \ref{fig:exclusionlimitsfk}, assuming $C_\Lambda = 0$ for simplicity. Possible improvements from LHC experiments regarding these bounds are discussed in section \ref{sec:operatorbounds}.

\paragraph{Monojet \& \texorpdfstring{$\slashed{E}_T$}{MET}:}
Both ATLAS and CMS have presented experimental searches with $8 \, \textrm{TeV}$ data for final states containing no leptons, one hard jet, missing transverse energy and at most a second slightly hard jet with $p_T > 30 \, \textrm{GeV}$ \cite{ATLAS:2012zim,CMS:rwa}. The ATLAS search defines four signal regions with both the $p_T$ of the leading jet and the $\slashed{E}_T$ to exceed $120, \, 220, \, 350, \, 500 \, \textrm{GeV}$, respectively. The CMS analysis, however, only defines signal regions in the missing transverse energy, which are $\slashed{E}_T > 250, \, 300, \, 350, \, 400, \, 450, \, 500, \, 550 \, \textrm{GeV}$, whilst requiring the leading jet $p_T > 110 \, \textrm{GeV}$. Additional suppression of QCD dijet background is handled by the two experiments in a different manner. ATLAS requires the azimuthal separation between the $\slashed{E}_T$ direction and the second leading jet, if present, to be greater than $0.5$. On the other hand, CMS only retains a two jet event if the azimuthal separation between the jets is less than $2.5$. In the absence of any deviation from the Standard Model, both experiments quote $95\%$ CL upper bounds on the signal visible cross section for all the signal regions defined above.

\begin{figure}[!ht]
	\begin{center}
		\includegraphics[scale=0.6]{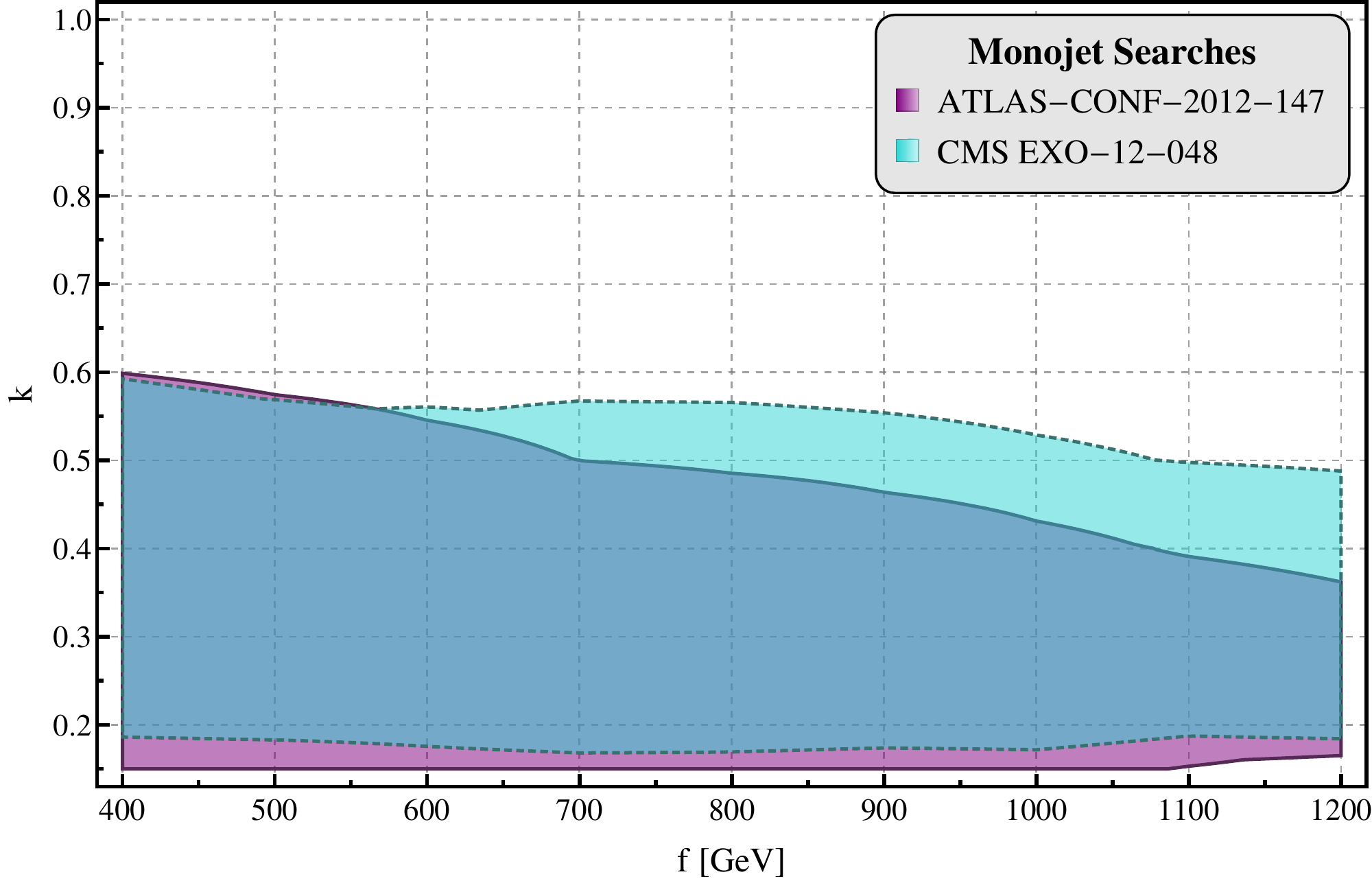} 
		\caption{$95\%$ CL exclusion limits from monojet \& $\slashed{E}_T$ direct searches at LHC8. The different contours represent the excluded regions from the latest monojet searches by ATLAS and CMS.}
		\label{fig:monojetmetexclusion}
	\end{center}
\end{figure}

Both monojet searches are suitable for final state topologies containing one or two hard jets and missing transverse energy. Hence, both LHT production modes $p \, p \to q_\subh \, q_\subh$ and $p \, p \to q_\subh \, A_\subh$ may contribute, provided the heavy quark partner decays to a quark and a heavy photon $q_\subh \to A_\subh \, q$. Therefore these searches have the highest exclusion power in the low-$k$ region ($0.2 \lesssim k \lesssim 0.6$). Indeed for $0.2 \lesssim k \lesssim 0.6$ the heavy quarks decay entirely into $A_\subh \, q$, giving the required final state topology. For higher values of $k$, the decays into heavy gauge bosons become kinematically allowed ($q_\subh \to W_\subh/Z_\subh \, q$), highly reducing the branching ratio $q_\subh \to A_\subh \, q$. In figure \ref{fig:monojetmetexclusion} one can observe the excluded contours by recasting both ATLAS and CMS monojet \& $\slashed{E}_T$ analysis.

\paragraph{Jets \& \texorpdfstring{$\slashed{E}_T$}{MET}:}
This category comprises all searches with at least two signal jets, missing transverse energy and no leptons in the final state. In the past searches of this kind have been studied in the context of the LHT model using Tevatron and early CMS data \cite{Carena:2006jx,Perelstein:2011ds}. In the last year, numerous searches interpreted in terms of supersymmetric final states have been presented by ATLAS and CMS for the $8 \, \textrm{TeV}$ data. All these searches have been analysed and the searches relevant for the LHT final states are by ATLAS \cite{ATLAS:2013fha,ATLAS:2013cma} and CMS \cite{Chatrchyan:2013lya}. The first ATLAS search is optimised for squarks and gluinos and the second for stops, whereas the CMS search looks more generally at squarks, sbottoms and gluinos. 

The ATLAS squark and gluino search \cite{ATLAS:2013fha} defines signal regions which require at least two, three, four, five or six jets, respectively. For those signal regions Standard Model backgrounds are reduced using cuts on $\Delta \phi$ between the jets and the missing transverse energy and stringent cuts on $\slashed{E}_T / m_\textrm{eff}$ and $m_\textrm{eff}$. In the LHT scenario these final states correspond to pair production of heavy gauge bosons and heavy quark partners or mixed states like $V_\subh \, q_\subh$, with subsequent decays $q_\subh \to V_{H} \, q$, $V_\subh \to V_{SM} \, A_\subh$, and all hadronic decays of the SM gauge bosons $V_{SM}$. 

\begin{figure}[!ht]
	\begin{center}
		\includegraphics[scale=0.6]{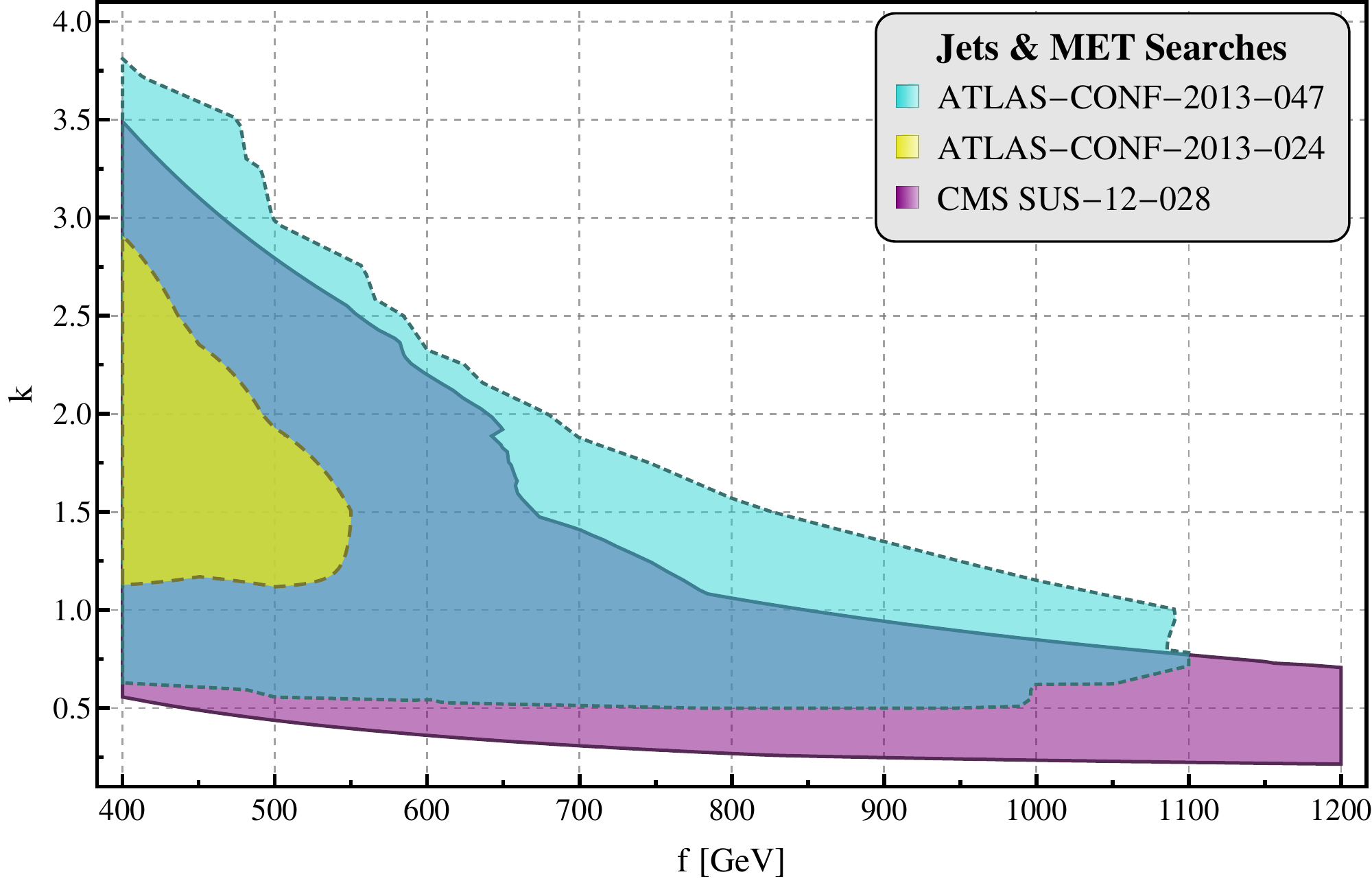} 
		\caption{$95\%$ CL exclusion limits from jets \& $\slashed{E}_T$ direct searches at LHC8. The different contours represent the excluded regions from the latest jets \& $\slashed{E}_T$ searches by ATLAS and CMS.}
		\label{fig:jetsmetexclusion}
	\end{center}
\end{figure}

The other ATLAS analysis \cite{ATLAS:2013cma} is optimised for searches of stop pair production, where each stop decays exclusively to hadronically decaying top and the LSP. The analysis requires at least six hard jets ($p_{T} > 35  \, \textrm{GeV}$), of which at least two must originate from bottom quarks. Three signal regions with significant missing transverse energy $\slashed{E}_T > 200, \, 300, \, 350 \, \textrm{GeV}$ are defined and two trijet systems should each roughly reconstruct the top mass. To further suppress the Standard Model background, cuts are placed on $\Delta \phi$ between the three highest-$p_T$ jets and $\slashed{E}_T$. On top of that the transverse mass $m_{T}$ between the $\slashed{E}_T$ and the b-tagged jet closest in $\Delta \phi$ to the $\slashed{E}_T$ direction is required to be greater than $175 \, \textrm{GeV}$. A possible LHT production mode is pair production of the T-odd top $t_\subh$ with subsequent stop like decay. Another production mode is two heavy quark partners where at least one of the quarks decays like $q_\subh \to Z_\subh \, q$, giving the required two $b$-jets.  

The CMS analysis \cite{Chatrchyan:2013lya} for squarks, including sbottoms, and gluinos looks at events with multiple jets, some of them $b$-jets, and significant missing energy. The analysis defines five signal regions tailored for the specific supersymmetry final states as $( N_\textrm{jet} , N_b ) = (2-3,0) ; (2-3, 1-2) ; (\geq 4, 1-2) ;  (\geq 4, 0) ;  (\geq 4, \geq 2)$. In order to suppress Standard Model background there are cuts on the transverse momenta of the jets and the scalar sum of the transverse momenta of the jets. Furthermore the $\alpha_T$ variable is used to protect against jet energy mismeasurement and is generalised to multi-jet final systems. The first signal region is a perfect fit for the production and decay $q_\subh \, q_\subh \to (A_\subh \, q) + (A_\subh \, q)$ which is efficient for low values of $k \lesssim 0.6$. The signal regions with at least four jets are instead efficient for the complementary region of $k \gtrsim 0.6$, where we can look at pair production of heavy gauge bosons and quark partners as well as the associated productions $V_\subh \, q_\subh$ with all hadronic final states, as already mentioned for the ATLAS analyses.

Each of the above searches provide $95\%$ CL upper bounds on the visible cross sections in the absence of any signal. The results from recasting the analysis are provided in figure \ref{fig:jetsmetexclusion}.

\paragraph{Leptons, jets \& \texorpdfstring{$\slashed{E}_T$}{MET}:}
In this section all searches involving leptons, at least two jets and missing transverse energy are considered, where some of the jets may be b-tagged. Indeed several searches exist by ATLAS and CMS, which match the latter final states. Here we only consider the constraining searches for the LHT model, in particular these are \cite{ATLAS:2012tna,ATLAS:2012maq,ATLAS:2013pla,ATLAS:2013tma}.

A search for supersymmetry using a single isolated lepton, at least four jets and missing transverse energy has been performed by ATLAS \cite{ATLAS:2012tna}. The lepton in the event can be either an electron or a muon, where both cases are considered separately and define a signal region each. Events with more than one lepton are vetoed. Each of the four jets in the event need to have $p_T > 80 \, \textrm{GeV}$ and additional kinematic cuts to suppress Standard Model background are: $\slashed{E}_T > 250 \, \textrm{GeV}$, $m_T (l, \slashed{E}_T) > 250 \, \textrm{GeV}$, $\slashed{E}_T/m_\textrm{eff} > 0.2$ and $m_\textrm{eff}^\textrm{inc} > 800 \, \textrm{GeV}$. In terms of LHT topologies, the production of two heavy quarks, which then decay to heavy gauge bosons $q_\subh \to W_\subh \, q$ or $\to Z_\subh \, q$ with at least one leptonically decaying $W$ at the end of the decay chain, results in exactly this final state. 

\begin{figure}[!ht]
	\begin{center}
		\includegraphics[scale=0.6]{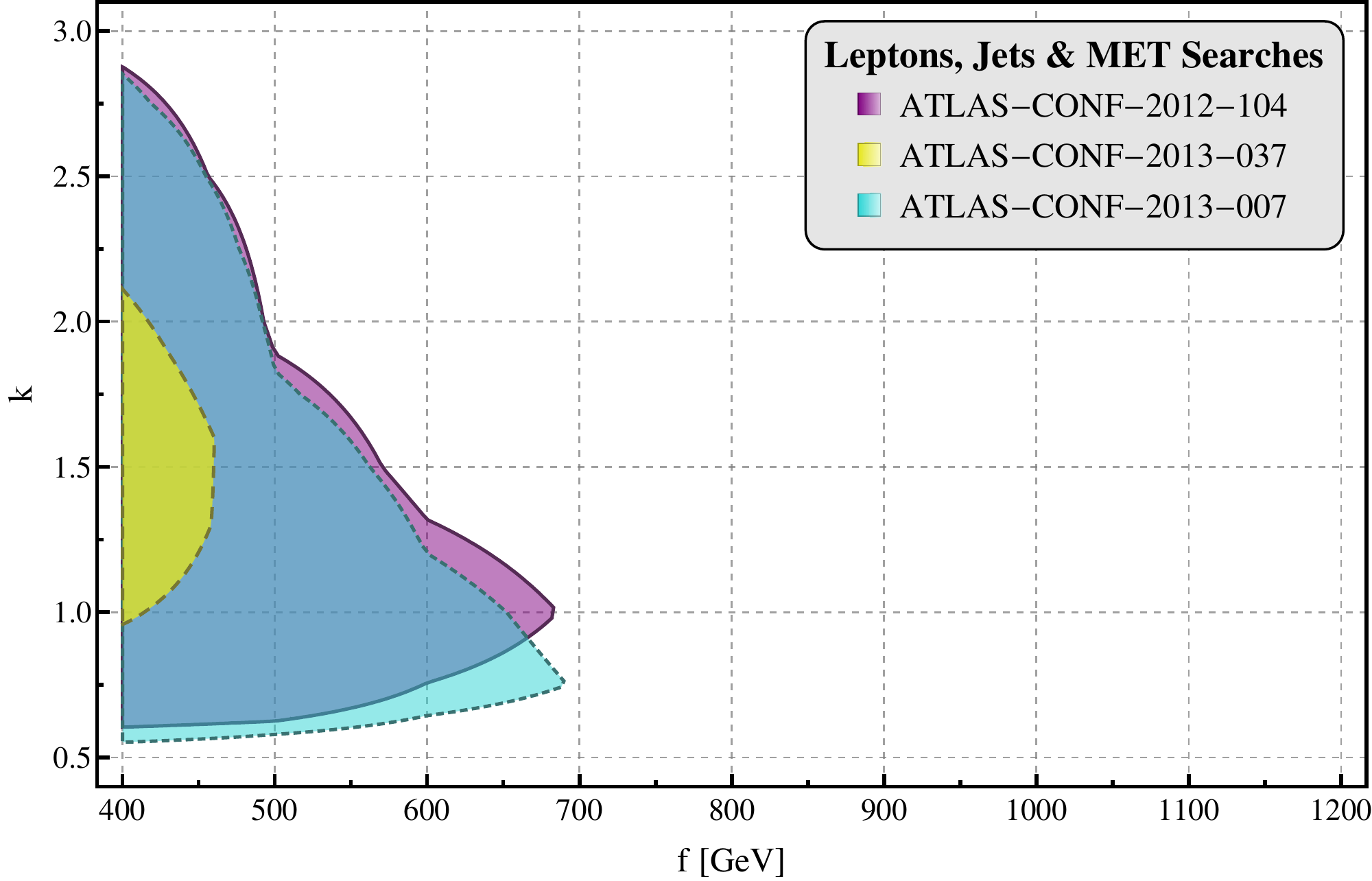} 
		\caption{$95\%$ CL exclusion limits from leptons, jets \& $\slashed{E}_T$ direct searches at LHC8.}
		\label{fig:leptonsjetsmetexclusion}
	\end{center}
\end{figure}

There is also a supersymmetry analysis focussing on stops by ATLAS \cite{ATLAS:2012maq}, which is updated in \cite{ATLAS:2013pla}. In this scenario the stop either decays into a top quark and the LSP or each top squark decays to a bottom quark and the lightest chargino. Therefore the analysis requires one isolated lepton, at least four jets of which at least one is b-tagged and significant missing transverse energy. Events with more than one lepton are vetoed and each of the signal regions implements various cuts used in supersymmetry searches like $\slashed{E}_T$, $m_\textrm{eff}$, $m_{T}$ and $m_{T2}$. Additionally for the signal regions involving top quarks, it is required to reconstruct the mass of the hadronically decaying top. The LHT production modes which may contribute are the same as before (pair production of heavy quark partners) with subsequent decays into gauge boson partners $W_\subh, \, Z_\subh$, but with semi-leptonic decays of the SM gauge bosons. As before, $b$-jets may arise from the decay 
of the Higgs boson from the $Z_\subh \to H \, A_\subh$ decay chain, or from the decay chain of the T-odd top partner $t_\subh \to t \, A_\subh$. 

The ATLAS search \cite{ATLAS:2013tma}, which was originally optimised for searches of gluino pair production, looks for two same sign leading leptons in combination with at least three jets and a significant amount of missing transverse energy. This search is divided into three signal regions with different number of jets and $b$-jets, but since the only final state in our model with two same sign leptons contains at most two additional jets, we only considered the first signal region. This signal region requires at least three jets: for our signal we rely on initial and final state QCD radiation for one additional jet, which is easily possible since the jets in this analysis only need to have $p_T > 40 \, \textrm{GeV}$. Further requirements on the event kinematics are: $\slashed{E}_T > 150 \, \textrm{GeV}$, $m_{T}(l,\slashed{E}_{T}) > 100 \, \textrm{GeV}$ and $m_\textrm{eff} > 400 \, \textrm{GeV}$. The only decay chain to achieve this final state is pair production of same charge quark partners $p p \to q_\subh q_\subh$ with subsequent decays into gauge boson partners $W_\subh$ with all leptonic decays for the $W$s. A similar analysis by CMS \cite{Chatrchyan:2012paa} is not efficient because it requires at least two b-tagged jets, for which there is no LHT process which matches this final state. 

From the searches described in this paragraph, $95\%$ CL exclusion limits in the $(f , k)$ plane can be extracted similar to the methods described before. The results from the recast are presented in figure \ref{fig:leptonsjetsmetexclusion}. From this we conclude that searches for both a single and two leptons perform similarly, as long as no $b$-jets in the final states are required.

\subsection{Combined exclusion limits}
\label{sec:combinedexclusionlimits}
It is interesting to compare (and finally combine) the results from electroweak precision physics, Higgs precision physics and direct searches at the LHC. 

\begin{figure}[!ht]
	\begin{center}
		\includegraphics[scale=0.6]{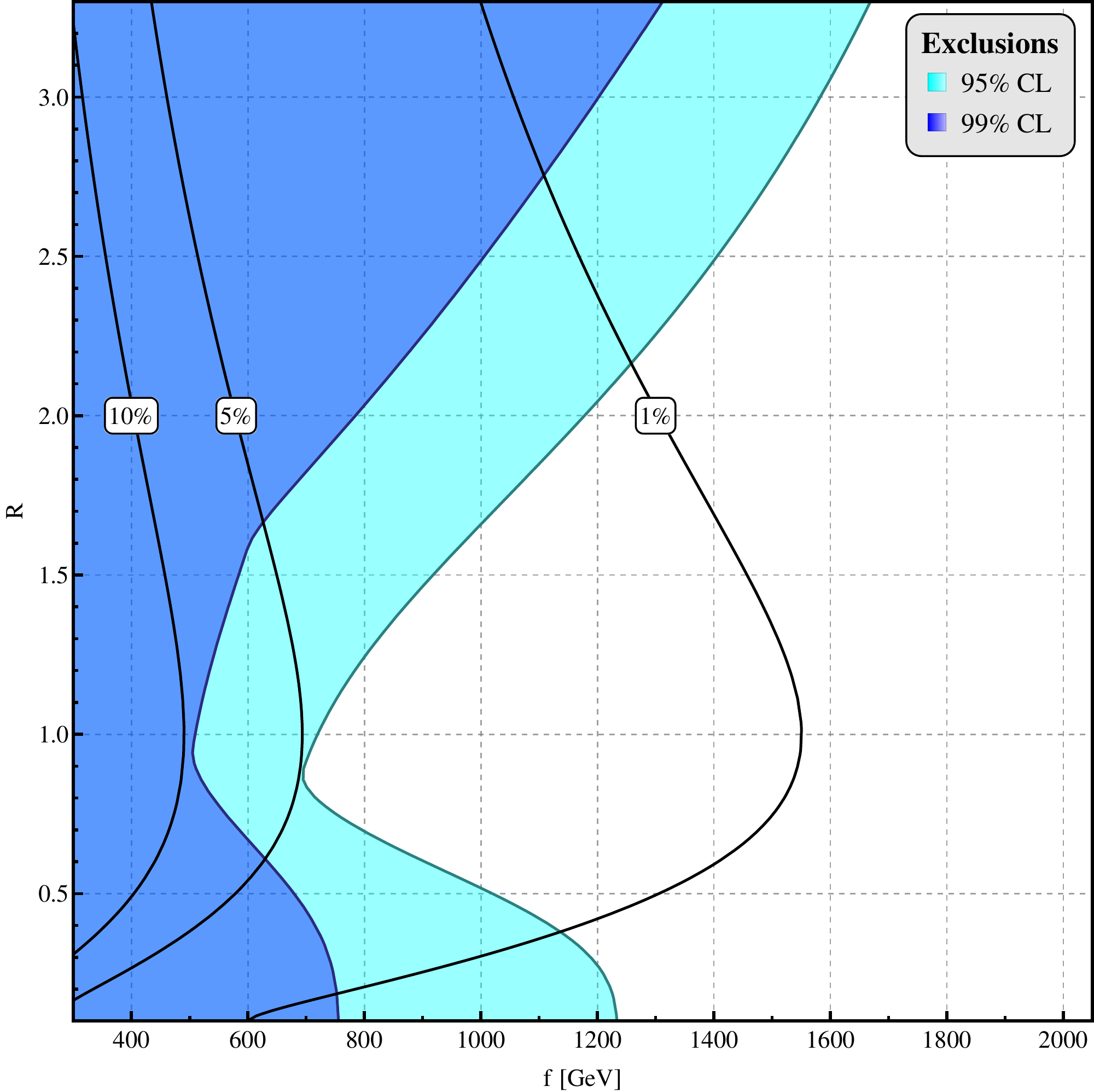} 
		\caption{Excluded parameter space regions at $95\%$ and $99\%$ CL from combination of EWPT and Higgs sector datasets. The thick black lines represent contours of required fine-tuning. The down-type Yukawa couplings are assumed to be from \emph{Case A}.}
		\label{fig:HiggsEWPT}
	\end{center}
\end{figure}

\paragraph{EWPO \& precision Higgs:}
By combining the $\chi^{2}$ analyses carried out separately for EWPT and the Higgs sector, as plotted in figure \ref{fig:HiggsEWPT}, the lower bound on the symmetry breaking scale is 
\begin{equation}
	\boxed{f \gtrsim \limitHiggsEWPT \, \textrm{GeV}} \quad \textrm{at} \quad 95\% \textrm{ CL}.
\end{equation}
By looking at this combination, the allowed fine-tuning is now worse than $10\%$, while still a small region in the parameter space could allow for a $\gtrsim 5\%$ fine-tuning. Results for the \emph{Case B} implementation of the down-type Yukawa couplings are provided in appendix \ref{sec:higgsewptcaseb}.

\begin{figure}[ht]
	\begin{center}
		\includegraphics[scale=0.6]{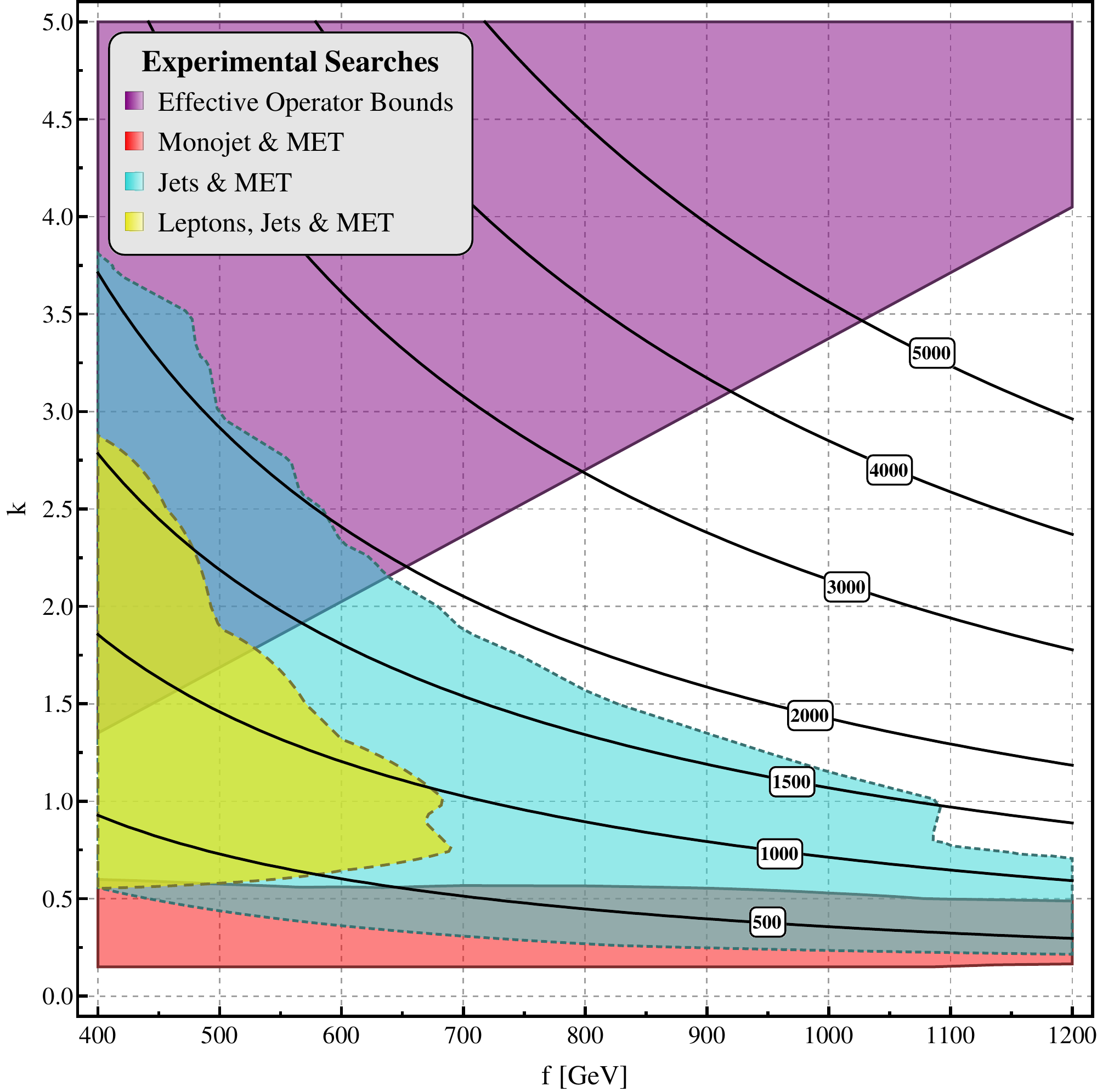}
		\caption{$95\%$ CL exclusion limits from direct searches at LHC8 displayed in the $(f,k)$ plane. The different categories comprise limits from operator bounds and searches from monojets, jets and leptons plus jets. The contour lines show the mass of the heavy quark partners.} 
		\label{fig:exclusionlimitsfk}
	\end{center}
\end{figure}

\paragraph{Direct LHC searches:}
Gathering all the exclusion limits from the aforementioned direct searches, the combined total exclusion limit in the $(f,k)$ plane is presented in figure \ref{fig:exclusionlimitsfk}. From these combined results the following lower bound on the symmetry breaking scale can be deduced:
\begin{equation}
	\boxed{f \gtrsim \limitDirectSearches \, \textrm{GeV}} \quad \textrm{at} \quad 95\% \textrm{ CL} .
\end{equation}
From the combination we can observe that the exclusion is dominated by all hadronic searches. Furthermore the requirement of b-jets or leptons in the final state only reduces the exclusion power for an LHT signal. This is mainly due to lower cross sections from reduced branching ratios for b-jets and leptons.

\section{Optimising current SUSY searches}
\label{sec:optimisingsusy}
The current dominant exclusion limits in the $(f,k)$ plane arise from three different direct searches: four fermion operator bounds, monojet plus missing energy and jets plus missing energy. In this section the possible optimisations for the direct searches and their effects on the parameter space of the LHT model are discussed. 

Using the tools for event generation as described before in the text, we generated event samples for some signal benchmark points in the $(f,k)$ parameter space and for background processes. We made sure to have significant statistics in order to obtain a reliable evaluation of the different cut efficiencies. We assumed a centre of mass energy of $8 \, \textrm{TeV}$. Following the set-up of the existing experimental analyses, we identified useful sets of kinematic cuts in order to reduce the backgrounds. To optimise the latter we then varied their values within sensible domains, evaluating the efficiencies for both signal and backgrounds as a function of the kinematic cuts. By reweighting the signal and background cross sections with the evaluated efficiencies, we obtained a map of the ratio $S/\sqrt{B}$, where S is the considered signal and B is the sum of all possible backgrounds, as a function of the cut values. By maximising the $S/\sqrt{B}$ ratio, we were then able to determine an optimised set of cut values which guarantees the highest exclusion (or discovery) power for the particular signal considered.

Once the optimised selection cuts are obtained, we multiplied the evaluated background efficiencies with the corresponding production cross sections. In this way we could determine the total number of expected background events, assuming an integrated luminosity as reported in the experimental papers. We then used a standard $CLs$ frequentist approach \cite{Cowan:2010js} to calculate the model independent $95 \%$ CL upper bound on the possible number of BSM signal events. In particular we calculated the $p$-values of the signal plus background and background only hypothesis, assuming Poisson probability for the number of observed events, and constructed a $CLs$ variable including systematic errors on the background. To retrieve the expected signal upper bound, the number of experimental observed events has been fixed to the number of expected events from the Standard Model. The upper bounds on the number of signal events can be finally translated into exclusion regions in the $(f,k)$ plane.

\subsection{Monojet \& \texorpdfstring{$\slashed{E}_T$}{MET} search}
\label{sec:monojetmetsearch}
Both the monojet searches by ATLAS \cite{ATLAS:2012zim} and CMS \cite{CMS:rwa} are designed to reach out into the more compressed supersymmetry spectra. In the LHT framework, too, the reach of these searches is in the more compressed part of parameter space, namely for lower values of $k$. This region is constrained by $m_{A_\subh} \leq m_{Q_\subh} \leq m_{V_\subh}$, which roughly implies $0.1 \lesssim k \lesssim 0.45$. In this region of parameter space the production modes $p \, p \to q_\subh \, q_\subh$ and $p \, p \to q_\subh \, A_\subh$ dominate in terms of production cross section, see table \ref{tab:lhtfinalstates}. These production modes lead to final states with one or two jets and significant amounts of missing transverse energy. This results from the fact that the heavy quarks uniquely decay as $q_\subh \to A_\subh \, q$. Even for higher values of $k$ the same final state is still a possibility: however, the branching ratio for this heavy quark decay rapidly decreases to $6-9$\%. Nevertheless we will still investigate the sensitivity of the monojet search also for higher values of $k \lesssim 1.0$. 

\begin{table}[!ht]
	\centering
	\begin{tabular}{l c c c c c}
		\toprule[1pt]
		Cut & BM1 & BM2 & BM3 & ATLAS & CMS \\
		\midrule[1pt]
		MET ($\slashed{E}_T$) & $170 \, \textrm{GeV}$ & $520 \, \textrm{GeV}$ & $370 \, \textrm{GeV}$ & $120$ & $250$ \\
		\midrule
		First jet $p_T$ & $120 \, \textrm{GeV}$ & $470 \, \textrm{GeV}$ & $250 \, \textrm{GeV}$ & $120$ & $110$ \\
		\midrule
		Second jet $p_T$ & $80 \, \textrm{GeV}$ & $310 \, \textrm{GeV}$ & $180 \, \textrm{GeV}$ & \xmark & \xmark \\
		\midrule
		Lepton veto & \cmark & \cmark & \cmark & \cmark & \cmark \\
		\midrule
		Two jet veto & \cmark & \cmark & \cmark & \cmark & \cmark \\
		\midrule
		$\Delta \phi (\slashed{E}_T, j_2) \geq$ & $0.5$ & $0.5$ & $0.5$ & $0.5$ & \xmark \\
		\midrule
		$\Delta \phi (j_1 , j_2)  \leq$ & $2.5$ & $2.5$ & $2.5$ & \xmark & $2.5$ \\
		\midrule[1pt]
		S$^{95}_{\textrm{exp}}$ & $1745$ & $8.4$ & $99.9$ & $45136$ & $3694$ \\
		\bottomrule[1pt]
	\end{tabular}
	\caption{Cut-flow table for the monojet \& $\slashed{E}_T$ optimization. In analogy with the existing analyses the lepton veto dismisses any event with an electron ($p_T > 10 \, \textrm{GeV}$), a muon ($p_T > 10 \, \textrm{GeV}$) or a tau ($p_T > 20 \, \textrm{GeV}$). The two jet veto removes all events with more than two jets satisfying $p_T > 30 \, \textrm{GeV}$ and $\eta < 4.5$. Shown in the first three columns are the optimised cuts for the chosen benchmark points. The last two columns show one of the signal regions of the ATLAS and CMS analysis \cite{ATLAS:2012zim,CMS:rwa}, for comparison and validation with the experimental results. S$^{95}_{\textrm{exp}}$ is the upper bound on the number of signal events obtained with the statistics method described at the beginning of section \ref{sec:optimisingsusy}.}
	\label{tab:monojetcutflow}
\end{table}

First the backgrounds and the ATLAS and CMS monojet analyses are discussed, then the procedure of optimising the kinematic cuts and finally potential exclusion contours in the $(f,k)$ plane are obtained. Since the two experimental analyses are based on different amounts of integrated luminosity, we decided to use a reference value of $20 \, \textrm{fb}^{-1}$ for the monojet proposal.

\paragraph{Backgrounds \& Analyses:}
The dominant backgrounds for monojet searches are $Z(\to \nu\nu) + \, \textrm{jets}$ and $W + \, \textrm{jets}$, with subleading contributions from $Z / \gamma^\ast (\to l^+ l^-) + \, \textrm{jets}$, multi-jet, $t \bar{t}$ and diboson ($WW$, $ZZ$, $WZ$) processes. All these processes have been simulated using the Monte Carlo chain described previously. The background samples have been generated applying the detector specifications reported in the ATLAS analysis paper.

Both the ATLAS and CMS analyses use roughly the same set of cuts to suppress the backgrounds. They share a lepton veto and a two jet veto, which forbids any final state with leptons or more than two hard jets. Furthermore they use cuts on $\Delta \phi$ between the missing energy and the second leading jet, and between the first and the second leading jet, by ATLAS and CMS, respectively. On top of these basic cuts, signal regions are defined which set varying cuts on missing transverse energy and the $p_T$ of the leading jet. The ATLAS search defines four signal regions with both the $p_T$ of the leading jet and the $\slashed{E}_T$ to exceed $120, \, 220, \, 350, \, 500 \, \textrm{GeV}$, respectively. The CMS analysis however only defines signal regions in the missing transverse energy, which are $\slashed{E}_T > 250, \, 300, \, 350, \, 400, \, 450, \, 500, \, 550 \, \textrm{GeV}$, whilst requiring the leading jet $p_T > 110 \, \textrm{GeV}$. Two reference ATLAS and CMS signal regions are shown in the last two columns of table \ref{tab:monojetcutflow}.

\begin{table}[!ht]
	\centering
	\begin{tabular}{c c c}
		\toprule[1pt]
		Benchmark & $f (\textrm{GeV})$ & $k$ \\
		\midrule[1pt]
		BM1 & $1600$ & $0.2$ \\
		BM2 & $2000$ & $0.4$ \\
		BM3 & $600$ & $0.8$ \\
		\bottomrule[1pt]
	\end{tabular}
	\qquad \qquad
	\begin{tabular}{l c}
		\toprule[1pt]
		Cut & Range \\
		\midrule[1pt]
		$\slashed{E}_T$ &  $[120, 600] \, \textrm{GeV}$  \\
		$p_T (j_{1})$ & $[100,600] \, \textrm{GeV}$ \\
		$p_T (j_{2})$ & $[0,450] \, \textrm{GeV}$ \\
		\bottomrule[1pt]
	\end{tabular}
	\caption{Benchmark scenarios (left) and ranges for the kinematic cuts (right) for the monojet proposal.}
	\label{tab:monojetbenchmarkandcuts}
\end{table}

\begin{figure}[ht]
	\begin{center}
		\includegraphics[scale=0.6]{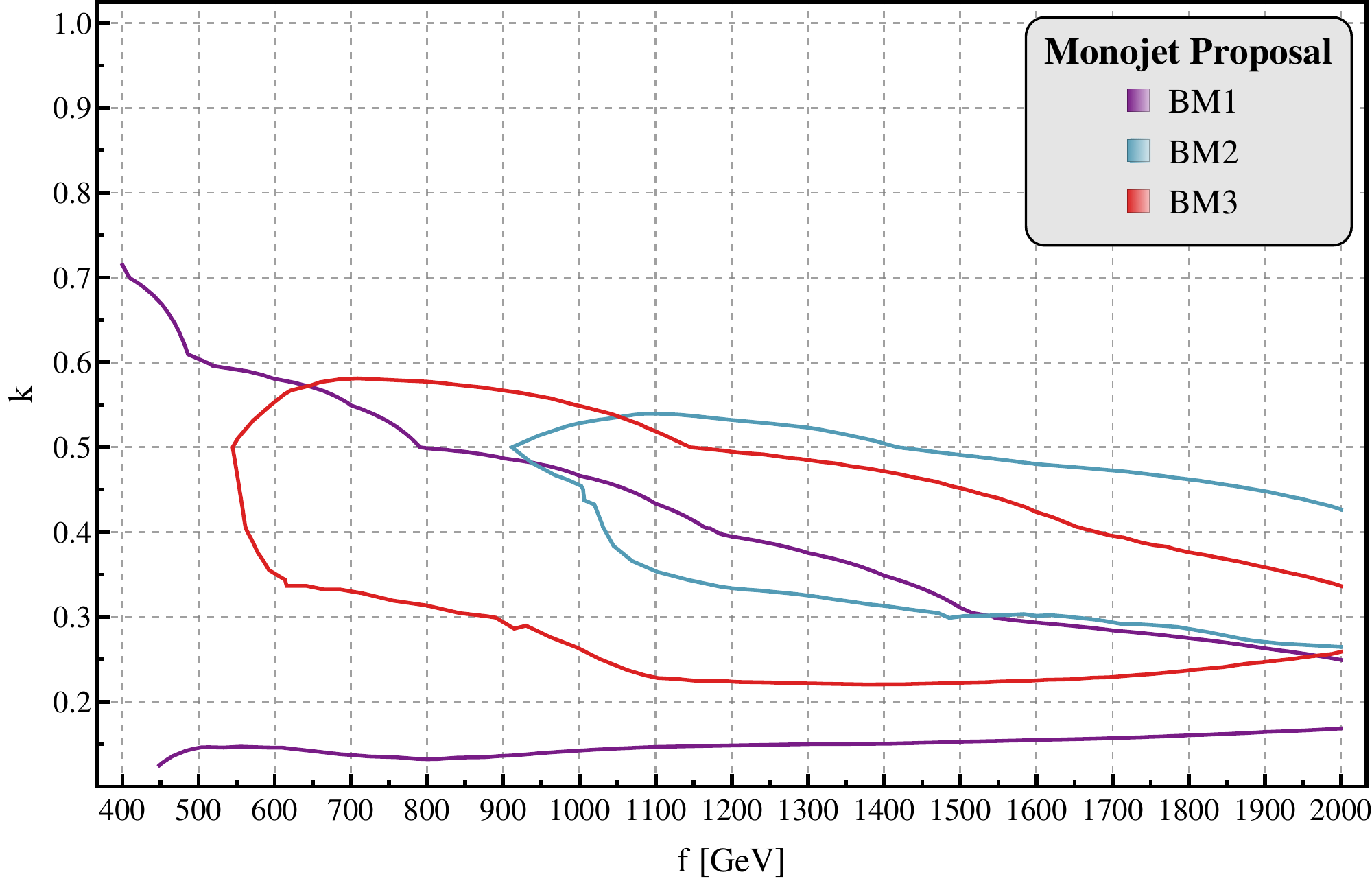}
		\caption{$95\%$ CL potential exclusion limits from the proposed optimised monojet search at LHC8 with $20 \, \textrm{fb}^{-1}$ displayed in the $(f,k)$ plane. The different contours correspond to different signal regions.} 
		\label{fig:proposallimitsmonojet}
	\end{center}
\end{figure}

\paragraph{Cut-flow:}
As discussed before, a set of kinematic cuts, including the ranges in which they are varied, is defined in table \ref{tab:monojetbenchmarkandcuts}. The minimum values for both the $\slashed{E}_T$ and $p_T$ cut are needed to allow for efficient triggering in both ATLAS and CMS detectors for the monojet searches. Then the optimal values for these cuts are determined for a set of benchmark scenarios, the latter listed in table \ref{tab:monojetbenchmarkandcuts} as well. In contrast to the monojet searches by ATLAS and CMS we do allow for a $p_T$ cut on the subleading jet, since our signal mainly consists of two jets. This topology has also been studied in \cite{Choudhury:2012xc}. Such a cut will aid significantly in the suppression of the background. For each of these benchmark points an optimal $S/\sqrt{B}$ is obtained for the values shown in table \ref{tab:monojetcutflow} and these cuts are then used to define three signal regions.

In general we observe that both the missing transverse energy cut and the cuts on the $p_T$ of the jets increase as the mass gap between the heavy quark and the heavy photon increases. This can be explained simply by the fact that the mass difference $m_{Q_\subh} - m_{A_\subh}$ will be translated to transverse momenta of both the jet and the heavy photon. Hence, the result will be high $p_T$ for the jets and a high missing transverse energy for high mass gaps.

\begin{figure}[ht]
	\begin{center}
		\includegraphics[scale=0.6]{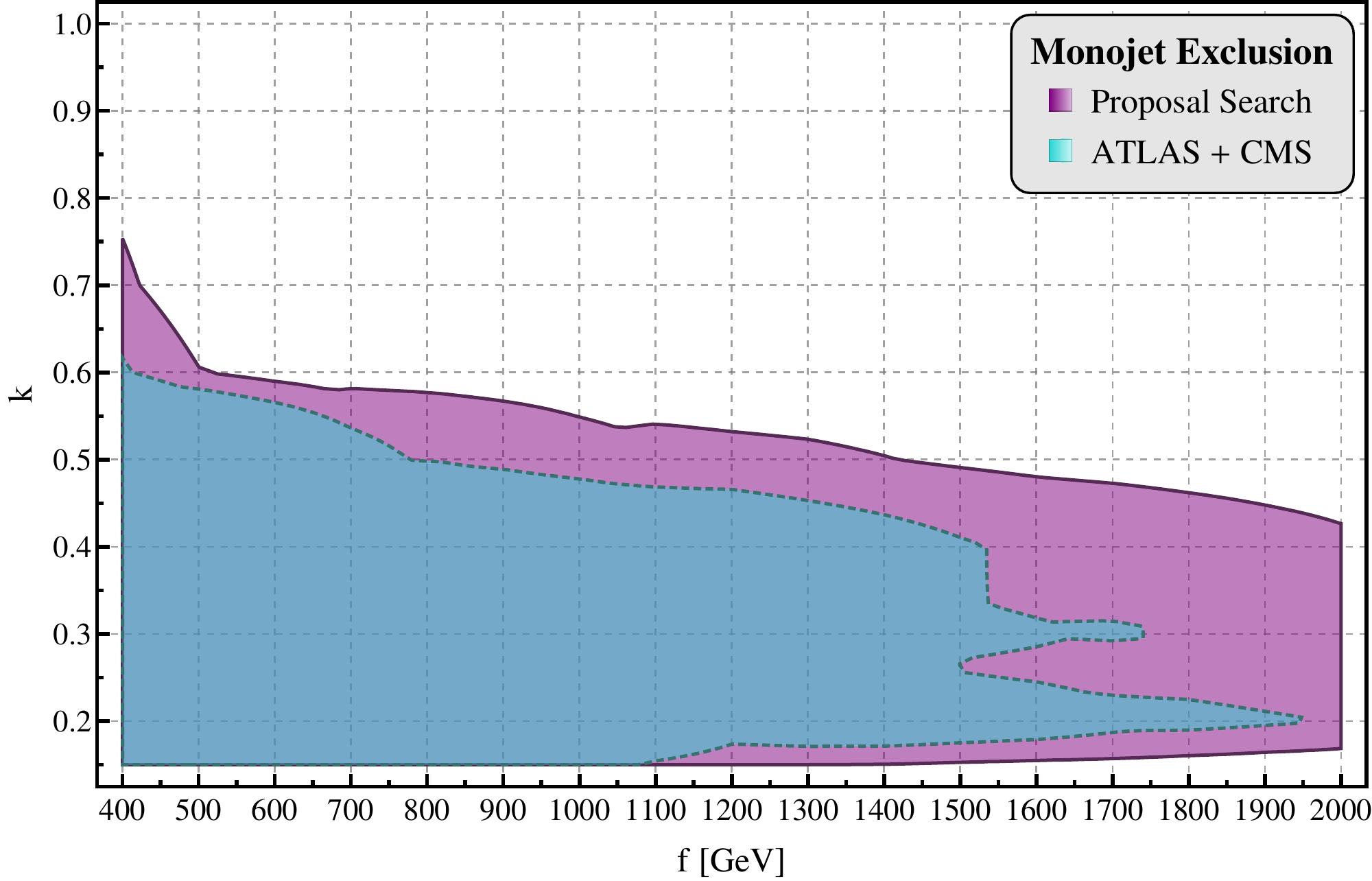}
		\caption{Comparison between the monojet proposal exclusion limits and the limits from the experimental signal regions. The upper bounds on the visible cross sections in both cases have been obtained using our statistics method. This is to ensure a comparison which does not depend on the background simulations nor the statistics method.} 
		\label{fig:proposallimitsmonojetcomparison}
	\end{center}
\end{figure}

\paragraph{Exclusion Limits:}
After having used the cut-flow procedure we essentially have an experimental monojet search with three signal regions corresponding to the three benchmark points. For each of the benchmark points we obtained the corresponding upper bounds on the visible cross sections, by dividing the signal upper bounds from table \ref{tab:monojetcutflow} by the luminosity. These limits can then be compared to the signal visible cross section as a function of $f$ and $k$ and exclusion contours can be drawn. The exclusion contours per signal region are provided in figure \ref{fig:proposallimitsmonojet}. 

The combined exclusion limit from all signal regions is given in figure \ref{fig:proposallimitsmonojetcomparison}, together with the result of the recasting procedure for the monojet analyses for comparison. Here we used the upper bound on the visible cross section evaluated with our statistics method for both the proposed signal regions as well as for the recasting of the experimental signal regions. This has been done in order to show that the increased exclusion power is genuinely due to the optimisation procedure. The results show that the range of the monojet proposal extends into the high $f$ regions for all $k$ values between $0.2$ and $0.6$. Above $k = 0.6$ the decay $q_\subh \to A_\subh \, q$ is too much suppressed and the cross sections are too low, whereas below $k = 0.2$ the spectrum is becoming compressed, reducing the proposal's sensitivity. We can conclude that there is room for improvement especially in the high $f$ regime, which will be vital in the future for excluding $f$ beyond a TeV.

\subsection{Jets \& \texorpdfstring{$\slashed{E}_T$}{MET} search}
\label{sec:jetsmetsearch}
The highest exclusion sensitivity in the LHT parameter space, as clearly pointed out in section \ref{sec:combinedexclusionlimits}, is achieved in jets \& $\slashed{E}_T$ final state topologies. This is mainly due to the higher available LHT signal cross section which could match the considered topology, see table \ref{tab:lhtfinalstates}. As described in section \ref{sec:directsearches} two available analyses scrutinising possible BSM signals in the jets \& $\slashed{E}_T$ final state are the ATLAS \cite{ATLAS:2013fha} and CMS \cite{Chatrchyan:2013lya}, where the former is optimised for searches of squarks and gluinos, while the latter more generally for searches of squarks, sbottoms and gluinos. The goal of this section is to reformulate the set-up of the ATLAS analysis assuming an LHT signal instead of a SUSY signal for which the analysis has been realised. In this way we will be able to propose an optimised set of selection cuts in order to reach the highest possible exclusion power for an LHT signal.

\paragraph{Experimental Analysis:}
We decided to focus on the ATLAS signal regions which require a hard $p_{T}$ cut on the third- and fourth-leading jet (SR B, C in ATLAS notation). The corresponding selection cuts used to reduce the backgrounds are summarised in table \ref{tab:ATLjetsMET}.

\begin{table}[!ht]
	\centering
	\begin{tabular}{l c | c | c | c}
		\toprule[1pt]
		Cut & SR BM & SR BT & SR CM & SR CT \\
		\midrule[1pt]
		Lepton veto & \multicolumn{4}{c}{\cmark} \\
		\midrule
		$n_{\text{jets}} \geq$ & \multicolumn{2}{c |}{3} & \multicolumn{2}{c}{4} \\
		\midrule
		$\slashed{E}_{T} >$ & \multicolumn{4}{c}{$160 \, \textrm{GeV}$} \\
		\midrule
		$p_T (j_{1}) >$ & \multicolumn{4}{c}{$130 \, \textrm{GeV}$} \\
		\midrule
		$p_T (j_{n}) >$ & \multicolumn{4}{c}{$60 \, \textrm{GeV}$} \\
		\midrule
		$\Delta \phi (j_{1,2,3} , \slashed{E}_T)_{\text{min}} >$ &  \multicolumn{4}{c}{0.4} \\
		\midrule
		$\slashed{E}_{T}/m_{\text{eff}}(n_{j}) >$ & 0.3 & 0.4 & \multicolumn{2}{c}{0.25}  \\
		\midrule
		$m_{\text{eff}}(\text{incl.}) >$ & $1.8 \, \textrm{TeV}$ & $2.2 \, \textrm{TeV}$ & $1.2 \, \textrm{TeV}$ & $2.2 \, \textrm{TeV}$ \\
		\bottomrule[1pt]
	\end{tabular}
	\caption{Selection cuts used in the ATLAS analysis \cite{ATLAS:2013fha} for the signal regions B (3j) and C (4j).}
	\label{tab:ATLjetsMET}
\end{table}

Notice that signal jets need to satisfy $p_{T}> 40 \; \textrm{GeV}$ and $|\eta|<2.8$, while signal electrons (muons) $p_{T}> 20 \; (10) \; \textrm{GeV}$ and $|\eta|<2.8 \; (2.4)$. The $\slashed{E}_{T}/m_{\text{eff}}(n_{j})$ cut in any $n$-jet channel uses a value of $m_{\text{eff}}$ constructed by the scalar sum of the transverse momenta only of the $n$ leading jets (and $\slashed{E}_{T}$), while the $m_{\text{eff}}(\text{incl.})$ selection includes all jets with $p_{T} > 40 \, \textrm{GeV}$ besides the $\slashed{E}_{T}$. For the cut on the minimal azimuthal separation between the $\slashed{E}_{T}$ direction and the reconstructed jets, only the three leading jets are considered. An additional requirement of $\Delta \phi (j , \slashed{E}_T) > 0.2$ is placed on all jets with $p_{T} > 40 \, \textrm{GeV}$.

\paragraph{Backgrounds:}
The dominant SM background processes are $W + \, \textrm{jets}$, $Z + \, \textrm{jets}$, top quark pairs, diboson, single top and multiple jets productions. The majority of the $W + \, \textrm{jets}$ background is composed by $W \to l \nu$ events in which no charged lepton is reconstructed, or $W \to \tau \nu$ with a hadronically decaying $\tau$. The largest part of the $Z + \, \textrm{jets}$ background comes from the $Z \to \nu \nu$  component, generating large $\slashed{E}_{T}$. Top quark single and pair production followed by semi-leptonic decays (both to a light charged lepton or to a $\tau$ lepton) can generate $\slashed{E}_{T}$, too, and pass the jet and lepton requirements at a non-negligible rate. The multi-jet background is caused by misreconstruction of jet energies in the calorimeters, leading to apparent $\slashed{E}_{T}$. The background samples have been generated applying the detector specifications reported in the analysis paper.

\paragraph{Cut-flow:}
As the signal events are regarded, we generated samples for three different choices of free parameters, with substantially different kinematical properties involved. These are summarised in table
\ref{tab:jetsmetbenchmarkandcuts}. The ranges of the kinematic cuts in which they are varied to obtain an optimal set-up are reported in table \ref{tab:jetsmetbenchmarkandcuts} as well.

\begin{table}[!ht]
	\centering
	\begin{tabular}{c c c}
		\toprule[1pt]
		Benchmark & $f (\textrm{GeV})$ & $k$ \\
		\midrule[1pt]
		BM1 & $600$ & $1.0$ \\
		BM2 & $700$ & $2.0$ \\
		BM3 & $1000$ & $1.0$ \\
		\bottomrule[1pt]
	\end{tabular}
	\quad \quad
	\begin{tabular}{l c}
		\toprule[1pt]
		Cut & Range \\
		\midrule[1pt]
		$n_{\text{jets}}$ & 3 or 4 \\
		$\slashed{E}_T$ &  $[100, 500] \, \textrm{GeV}$  \\
		$p_T (j_{1})$ & $[100,400] \, \textrm{GeV}$ \\
		$p_T (j_{n})$ & $[40,100] \, \textrm{GeV}$ \\
		$m_{\text{eff}}(\text{incl.})$ & $[1.2, 3.0] \, \textrm{TeV}$ \\
		\bottomrule[1pt]
	\end{tabular}
	\caption{Benchmark scenarios (left) and ranges for the kinematic cuts (right) for the jets \& $\slashed{E}_{T}$ proposal.}
	\label{tab:jetsmetbenchmarkandcuts}
\end{table}

The lepton veto and an additional cut of $\Delta \phi (j_{1,2,3} , \slashed{E}_T)_{\text{min}} > 0.4$ are applied in each signal region, in order to further reduce the different backgrounds. Clearly, if a cut on the $p_{T}$ on the $n$-th leading jet is applied ($p_T (j_{n})$), at least $n$ signal jets are required to be present in the final state. For each benchmark point, an optimal $S/\sqrt{B}$ ratio is obtained for the values shown in table \ref{tab:jetsmetcutflow}. 

\begin{table}[!ht]
	\centering
	\begin{tabular}{l c | c | c | c | c | c}
		\toprule[1pt]
		Cut & BM1$_{3j}$ & BM2$_{3j}$ & BM3$_{3j}$ & BM1$_{4j}$ & BM2$_{4j}$ & BM3$_{4j}$   \\
		\midrule[1pt]
		Lepton veto & \multicolumn{6}{c }{\cmark} \\
		\midrule
		$n_{jets}$ & \multicolumn{3}{c |}{3}  & \multicolumn{3}{c}{4} \\
		\midrule
		$\slashed{E}_T$ & $200 \, \textrm{GeV}$ & $340 \, \textrm{GeV}$ & $400 \, \textrm{GeV}$ & $200 \, \textrm{GeV}$ & $300 \, \textrm{GeV}$ & $400 \, \textrm{GeV}$ \\
		\midrule
		$p_T (j_{1})$  & $120 \, \textrm{GeV}$ & $380 \, \textrm{GeV}$ & $180 \, \textrm{GeV}$ & $140 \, \textrm{GeV}$ & $320 \, \textrm{GeV}$ & $180 \, \textrm{GeV}$ \\
		\midrule
		$p_T (j_{n})$ & $100 \, \textrm{GeV}$ & $100 \, \textrm{GeV}$ & $100 \, \textrm{GeV}$ & $70 \, \textrm{GeV}$ & $80 \, \textrm{GeV}$ & $100 \, \textrm{GeV}$ \\
		\midrule
		$\Delta \phi (j_{1,2,3} , \slashed{E}_T)_{\text{min}}$ & \multicolumn{6}{c }{0.4} \\
		\midrule
		$m_{\text{eff}}(\text{incl.})$ & $1.2 \, \textrm{TeV}$ & $2.8 \, \textrm{TeV}$ & $2.1 \, \textrm{TeV}$ & $1.2 \, \textrm{TeV}$ & $2.6 \, \textrm{TeV}$ & $2.1 \, \textrm{TeV}$ \\
		\midrule[1pt]
		S$^{95}_{\textrm{exp}}$ & $298$ & $3.5$ & $11.3$ & $154$ & $3.5$ & $4.2$ \\
		\bottomrule[1pt]
	\end{tabular}
	\caption{Cut-flow table for the jets \& $\slashed{E}_T$ optimization. In analogy with the existing analysis the lepton veto dismisses any event with an electron (muon) with $p_{T}> 20 \; (10) \; \textrm{GeV}$ and $|\eta|<2.8 \; (2.4)$. S$^{95}_{\textrm{exp}}$ is the upper bound on the number of signal events obtained with the statistics method described at the beginning of section \ref{sec:optimisingsusy}.}
	\label{tab:jetsmetcutflow}
\end{table}

A few general observations can be made. First of all, the required cut on the effective mass ($m_{\text{eff}}$) increases with both $f$ and $k$: this is indeed a consequence of the increasing mass splitting between the \emph{mother} and \emph{daughter} particles in the decay chain, namely the heavy quark $q_\subh$ and the heavy photon $A_\subh$, respectively. If one considers the (light) quarks as massless, the effective mass in the heavy quark pair production could be indeed approximated with $m_{\text{eff}} \sim 2(m_{q_\subh} - m_{A_\subh})$. A second observation is that the required $\slashed{E}_T$ cut grows with $f$, namely again with the mass difference between $q_\subh$ and $A_\subh$. From the previous observations it clearly follows that if a hard cut on $m_{\text{eff}}$ is required together with a milder cut on $\slashed{E}_T$, at least a very hard jet is required in the spectrum: this is indeed the case for the benchmark points with relatively low values of $f$ and higher values of $k$. 

Compared to the values of table \ref{tab:ATLjetsMET}, one can see that an increased exclusion power  could generically be gained by increasing the values of the cuts on the effective mass $m_{\text{eff}}$ and on the missing energy $\slashed{E}_{T}$, especially for regions in the parameter space with higher values of $f$ and $k$.

\paragraph{Exclusion Limits:}
Assuming an experimental search with the set-up summarised in table \ref{tab:jetsmetcutflow}, we evaluated for each signal region the upper bound on the number of BSM signal events, under the hypothesis of exact overlap between background expectation and experimental yield, as described at the beginning of section \ref{sec:optimisingsusy}.

This gave us the opportunity to validate our methods, namely the reliability of our background samples, the recasting procedure and of the statistics method. By applying the set-up of the original ATLAS analysis on our background samples, we were able indeed to compare the expected number of background events with the reported numbers in the experimental paper, as well as the expected upper bounds on BSM events. The result of this comparison is summarised in table \ref{tab:validation}: the results are clearly consistent within the reported uncertainties.

\begin{table}[!ht]
	\centering
	\begin{tabular}{ c | c | c | c | c | c }
		\toprule[1pt]
		 & SR BM & SR BT & SR CM & SR CT & SR D \\
		\midrule[1pt]
		\multicolumn{6}{c }{ATLAS analysis \cite{ATLAS:2013fha}} \\ 
		\midrule
		Total bkg & $33 \pm 7$ & $2.4 \pm 1.4$ & $210 \pm 40$ & $1.6 \pm 1.4$ & $15 \pm 5$ \\
		S$^{95}_{\text{exp}}$ & $17.0^{+6.6}_{-4.6}$ & $5.8^{+2.9}_{-1.8}$ & $72.9^{+23.6}_{-18.0}$ & $3.3^{+2.1}_{-1.2}$ & $13.6^{+5.1}_{-3.5}$ \\ 
		\midrule[1pt]
		\multicolumn{6}{c }{Recasting procedure} \\ 
		\midrule
		Total bkg & $30.2 \pm 9.1$ & $3.2 \pm 1.6$ & $218.5 \pm 43.7$ & $2.4 \pm 1.2$ & $15.2 \pm 4.5$ \\
		S$^{95}_{\text{exp}}$ & $21.0$ & $5.4$ & $90.2$ & $4.3$ & $12.2$ \\
		\bottomrule[1pt]
	\end{tabular}
	\caption{Procedure validation: comparison between reported experimental results \cite{ATLAS:2013fha} and recasting procedure. In particular, the total number of background events and the corresponding $95 \%$ CL expected upper bound on BSM signal events (S$^{95}_{\text{exp}}$) are shown.}
	\label{tab:validation}
\end{table}

\begin{figure}[ht]
	\begin{center}
		\includegraphics[scale=0.6]{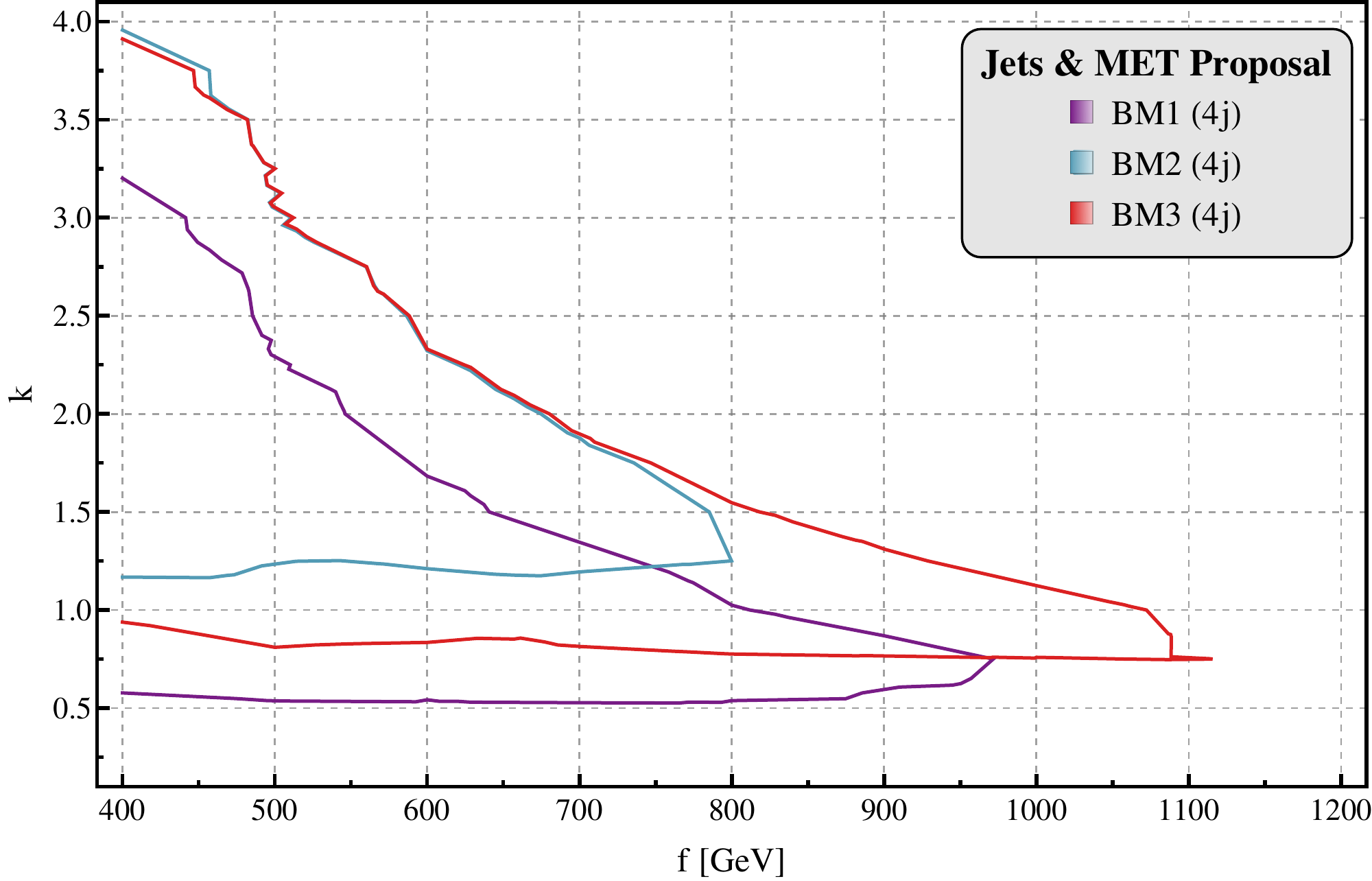}
		\caption{$95\%$ CL potential exclusion limits from the proposed optimised jets \& $\slashed{E}_T$ search at LHC8 with $20.3 \, \textrm{fb}^{-1}$ displayed in the $(f,k)$ plane. The different contours correspond to different signal regions.} 
		\label{fig:proposallimitsjetsmet}
	\end{center}
\end{figure}

The upper bounds on the visible cross section within the optimised signal regions can be extracted from the last row of table \ref{tab:jetsmetcutflow}. These limits can then be compared to the LHT signal visible cross section as a function of $f$ and $k$, and exclusion contours can be drawn. In particular, the exclusion contours per signal region are reported in figure \ref{fig:proposallimitsjetsmet}. It should be noticed that only the signal regions requiring at least four jets in the final state are included in the latter plot, since it turned out that they possess higher exclusion power than the corresponding signal regions which require at least three jets. 

\begin{figure}[ht]
	\begin{center}
		\includegraphics[scale=0.6]{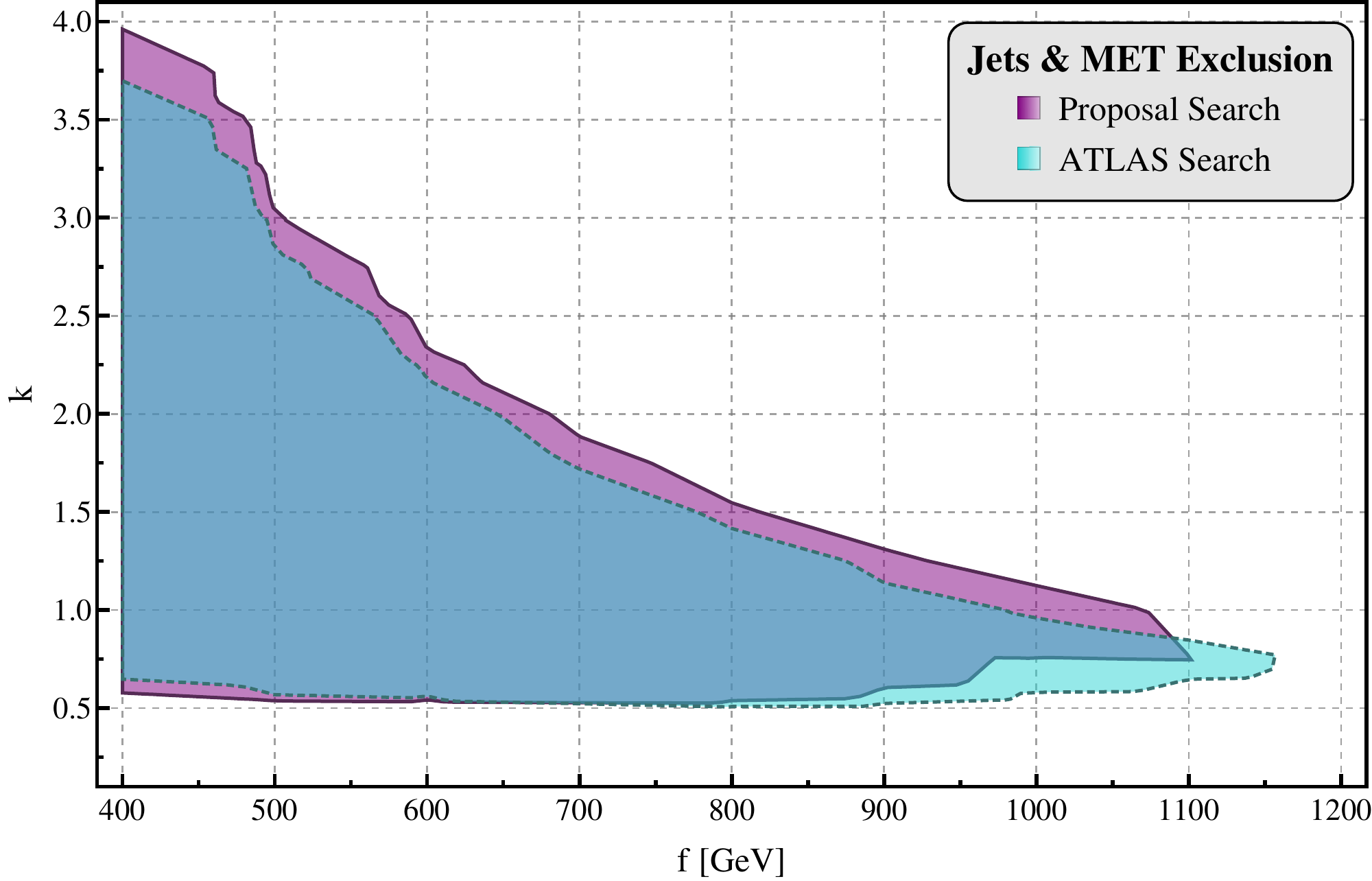}
		\caption{Comparison between the jets \& $\slashed{E}_T$ proposal exclusion limits and the limits from the experimental signal regions. The upper bounds on the visible cross sections in both cases have been obtained using our statistics method. This to ensure a comparison which does not depend on the background simulations nor the statistics method.} 
		\label{fig:proposallimitsjetsmetcomparison}
	\end{center}
\end{figure}

In figure \ref{fig:proposallimitsjetsmetcomparison} the combined exclusion limits from all signal regions are drawn, together with the result of the recasting procedure of the ATLAS analysis. It is to be noted that in the latter plot we used the upper bound on the visible cross section evaluated with our statistics method for both the proposed signal regions as well as for the recasting of the ATLAS signal regions. From figure \ref{fig:proposallimitsjetsmetcomparison} we can see that there is only small room for improvement in the jets \& $\slashed{E}_{T}$ final state topology, if one relies only on the set-up of the existing experimental searches. The improvement of the exclusion in the $f$-direction can be estimated to roughly $50 \, \textrm{GeV}$ for fixed value of $k$.

\subsection{Operator bounds}
\label{sec:operatorbounds}
As an aside to supersymmetric searches also operator bounds are important. Although the operator bounds on four fermion operators come with an intrinsic uncertainty from non-perturbative physics above the scale $\Lambda$, they provide both a viable and crucial method to constrain LHT parameter space. The peculiarity of the box diagrams generating four fermion contact interactions \cite{Hubisz:2005tx}, is that they provide an upper bound for $k$ given a scale $f$. On the other hand, the direct searches rather give a lower bound on $k$, hence the interplay between both allows to constrain the LHT model in the $(f,k)$ plane. From equation \eqref{eq:kfbound} one can immediately see that increasing the scale of non-flavour violating four fermion operators beyond the $25 \, \textrm{TeV}$ range will exclude even larger portions of parameter space for even lower values of $k$. We do expect LHC to improve these bounds, since at the moment only $7 \, \textrm{TeV}$ analyses for operator bounds of this form are available \cite{ATLAS:2012pu,Chatrchyan:2012bf}. The $8 \, \textrm{TeV}$ results on four-quark operators are already expected to become competitive with the LEP four-lepton operators, and we do expect the $14 \, \textrm{TeV}$ results to be really constraining for LHT parameter space.

\section{Conclusion}
\label{sec:conclusion}
In this work the Littlest Higgs with T-parity has been discussed in the threefold context of electroweak precision physics, Higgs precision physics and direct LHC searches, combining constraints from all possible corners of high energy physics. For this purpose an up-to-date overview of the relevant phenomenology for direct searches at the LHC has been presented. This has been used to discuss possible topologies which could mimic supersymmetry searches and therefore be constraining for the LHT parameter space. With this knowledge available we undertook the endeavour of constraining the LHT model with the most recent $8 \, \textrm{TeV}$ LHC data from ATLAS and CMS.

In principle, one wants to get the highest amount of information available on the parameter space of the LHT model, but the main goal is clearly to obtain the most stringent limit on the symmetry breaking scale $f$ in this model. This scale is the main parameter of the model as it sets the absolute scale for the whole symmetry breaking pattern and gives the connection to its strongly interacting UV completion. As it is also intimately connected to the amount of fine-tuning in the model, it provides a measure of the naturalness of the model. From the combination of EWPT and Higgs precision physics, we derive a lower limit on the symmetry breaking scale $f$ of 
\begin{equation*}
	\boxed{f \gtrsim \limitHiggsEWPT \, \textrm{GeV}} \quad \textrm{at} \quad 95\% \textrm{ CL} .
\end{equation*}
On the other hand, from direct searches the lower limit reads 
\begin{equation*}
	\boxed{f \gtrsim \limitDirectSearches \, \textrm{GeV}} \quad \textrm{at} \quad 95\% \textrm{ CL} .
\end{equation*}
With this result in mind, the implementation of the Littlest Higgs with T-parity is still natural, since the tuning is only roughly of the order of $5\%$. Note that direct searches are by now becoming competitive with constraints from indirect measurements like EWPT and Higgs precision physics.

The real potential for discovering and constraining the LHT model has been analysed as well. The exclusion possibilities from the monojet \& $\slashed{E}_T$ and the jets \& $\slashed{E}_T$ searches have been optimised for the LHT model. We performed an exhaustive scan over both the parameters $f$ and $k$ --- the coupling in the mirror fermion sector --- as well as the relevant kinematic cuts to analyse the exclusion potential of the $8 \, \textrm{TeV}$ run with roughly $20 \, \textrm{fb}^{-1}$. The results show that current direct searches can become competitive with indirect searches, though would not be able to push the exclusion limits much beyond. However, direct searches can cover interesting regions of the parameter space, which are left untouched by indirect constraints. In conclusion, the Little Higgs model with T-parity will hold its natural status during the LHC8 era.

Most importantly we would like to stress that we presented a consistent method to constrain the Littlest Higgs model with T-parity using direct searches. Even though at the moment direct searches are less constraining than indirect methods, these form a more direct and therefore more robust method to constrain the LHT parameter space. Improvements in four-fermion operator bounds, as well as the optimised direct searches, can be used by the CMS and ATLAS experiments to either discover or falsify the natural LHT model with the $14 \, \textrm{TeV}$ run. We therefore hope that the collaborations will extend the kinematic regime of their simplified model searches for supersymmetry, since we have shown that a recasting procedure provides a powerful method in constraining the LHT parameter space. 

\newpage

\appendix

\section{LHT topologies}
\label{sec:lhttopologies}

\begin{table}[!ht]
	\centering
	\begin{tabular}{c c c | c | c c | c c}
		\toprule[1pt]
		\multicolumn{3}{c|}{final state} & \multirow{2}{*}{\begin{tabular}{c} \vspace{1.5mm} production \\ modes \end{tabular}} & \multicolumn{2}{c}{$\sigma_{8 \, \textrm{TeV}} \times \textrm{Br}$ $(\textrm{fb})$} & \multicolumn{2}{c}{$\sigma_{14 \, \textrm{TeV}} \times \textrm{Br}$ $(\textrm{fb})$} \\
		\cmidrule{1-3} \cmidrule{5-6} \cmidrule{7-8}
		$\# \, l^\pm$ & \# jets & $\slashed{E}_T$ & & $k=1.0$ & $k=0.4$ & $k=1.0$ & $k=0.4$ \\ 
		\midrule[1pt]
		\multirow{1}{*}{$0$} & \multirow{1}{*}{$1$} & \multirow{1}{*}{\cmark} & $q_\subh A_\subh$ & $0.24$ & $1.1 \! \times \! 10^2$ & $2.1$ & $4.5 \! \times \! 10^2$ \\
		\midrule
		\multirow{1}{*}{$0$} & \multirow{1}{*}{$2$} & \multirow{1}{*}{\cmark} & $q_\subh q_\subh$ & $0.56$ & $5.6 \! \times \! 10^3$ & $5.2$ & $3.2 \! \times \! 10^4$ \\
		\midrule
		\multirow{2}{*}{$0$} & \multirow{2}{*}{$3$} & \multirow{2}{*}{\cmark} & $q_\subh W^\pm_\subh$ & $0.73$ & $14$ & $8.0$ & $77$ \\
		& & & $q_\subh Z_\subh$ & $0.76$ & $8.6$ & $8.0$ & $49$ \\
		\midrule
		\multirow{4}{*}{$0$} & \multirow{4}{*}{$4$} & \multirow{4}{*}{\cmark} & $q_\subh q_\subh$ & $4.0$ & $9.1 \! \times \! 10^2$ & $35$ & $5.6 \! \times \! 10^3$ \\
		& & & $W^\pm_\subh W^\mp_\subh$ & $1.9$ & low & $9.1$ & low \\
		& & & $W^\pm_\subh Z_\subh$ & $4.8$ & low & $23$ & low \\
		& & & $Z_\subh Z_\subh$ & $0.56$ & low & $3.0$ & low \\
		\midrule
		\multirow{1}{*}{$0$} & \multirow{1}{*}{$4$} & \multirow{1}{*}{\xmark} & $T^+ q$ & $2.0$ & $2.0$ & $17$ & $17$ \\
		\midrule
		\multirow{2}{*}{$0$} & \multirow{2}{*}{$5$} & \multirow{2}{*}{\cmark} & $q_\subh W^\pm_\subh$ & $5.1$ & \xmark & $54$ & \xmark \\
		& & & $q_\subh Z_\subh$ & $4.1$ & \xmark & $44$ & \xmark \\
		\midrule
		\multirow{2}{*}{$0$} & \multirow{2}{*}{$6$} & \multirow{2}{*}{\cmark} & $q_\subh q_\subh$ & $1.6$ & $9.7 \! \times \! 10^2$ & $1.7 \! \times \! 10^2$ & $6.0 \! \times \! 10^3$ \\
		& & & $T^- T^-$ & $2.5$ & $2.5$ & $25$ & $25$ \\
		\midrule
		\multirow{4}{*}{$l^\pm$} & \multirow{4}{*}{$2$} & \multirow{4}{*}{\cmark} & $q_\subh q_\subh$ & $0.058$ & $9.0 \! \times \! 10^2$ & $1.1$ & $5.6 \! \times \! 10^3$ \\
		& & & $W^\pm_\subh W^\mp_\subh$ & $0.77$ & low & $3.9$ & low \\
		& & & $W^\pm_\subh Z_\subh$ & $2.1$ & low & $10$ & low \\
		& & & $T^+ q$ & $1.3$ & $1.2$ & $10$ & $10$ \\
		\midrule
		\multirow{2}{*}{$l^\pm$} & \multirow{2}{*}{$3$} & \multirow{2}{*}{\cmark} & $q_\subh W^\pm_\subh$ & $3.5$ & \xmark & $37$ & \xmark \\
		& & & $q_\subh Z_\subh$ & $0.99$ & \xmark & $11$ & \xmark \\
		\midrule
		\multirow{2}{*}{$l^\pm$} & \multirow{2}{*}{$4$} & \multirow{2}{*}{\cmark} & $q_\subh q_\subh$ & $7.4$ & $9.7 \! \times \! 10^2$ & $82$ & $6.0 \! \times \! 10^3$ \\
		& & & $T^- T^-$ & $2.2$ & $2.2$ & $21$ & $21$ \\
		\midrule
		\multirow{1}{*}{$l^+l^-$} & \multirow{1}{*}{$0$} & \multirow{1}{*}{\cmark} & $W^\pm_\subh W^\mp_\subh$ & $0.32$ & low & $1.7$ & low \\
		\midrule
		\multirow{1}{*}{$l^+l^-$} & \multirow{1}{*}{$1$} & \multirow{1}{*}{\cmark} & $q_\subh W^\pm_\subh$ & $0.54$ & \xmark & $5.8$ & \xmark \\
		\midrule
		\multirow{2}{*}{$l^+l^-$} & \multirow{2}{*}{$2$} & \multirow{2}{*}{\cmark} & $q_\subh q_\subh$ & $1.1$ & \xmark & $11$ & \xmark \\
		& & & $T^- T^-$ & $0.47$ & $0.47$ & $4.6$ & $4.6$ \\
		\midrule
		\multirow{1}{*}{$l^\pm l^\pm$} & \multirow{1}{*}{$2$} & \multirow{1}{*}{\cmark} & $q_\subh q_\subh$ & $0.37$ & \xmark & $2.7$ & \xmark \\
		\bottomrule[1pt]
	\end{tabular}
	\caption{Overview of the relevant final states for LHC8 experimental searches in LHT models. The final states are classified according to the number of leptons and jets and whether they contain missing energy, and all the production modes contributing to each final state are listed. Note that the cross sections depend on $f$ and $k$, and also $R$ if the mode involves $T^\pm$. The last columns contain $\sigma \times \textrm{Br}$ for each of the production modes for fixed $f=750 \, \textrm{GeV}$ and $R=1.0$. A \xmark ~ indicates a mode without available phase space, whereas \emph{low} indicates negligible cross section at the LHC.}
	\label{tab:lhtfinalstates}
\end{table}

\section{Higgs precision data}
\label{sec:higgsdata}

\begin{table}[!ht]
	\centering
	\renewcommand{\arraystretch}{1.1}
	\begin{tabular}{ l | c c | c c | c }
		\toprule[1pt]
		Channel & $\hat \mu$ $(7 \, \textrm{TeV})$ & $\zeta_i^{\rm (g, V, t)}$ (\%)&  $\hat \mu$ $(8 \, \textrm{TeV})$ & $\zeta_i^{\rm (g,V,t)}$ (\%) & Refs. \\
		\midrule[1pt]
		$b \bar b$ (VH) & combination & --- & ${-0.42^{+1.05}_{-1.05}}$ & $(0,100,0)$ & \cite{ATLAS:NOINSP:bb} \\
		$b \bar b$ ($ttH$)& $3.81 \pm 5.78$ & $(0,30,70)$ & ---  & --- & \cite{ATLAS:2012cpa} \\
		\midrule
		$\tau \tau$ & combination & --- & $0.7^{+0.7}_{-0.7}$ & $(20,80,0)$& \cite{ATLAS:2012dsy} \\
		\midrule
		$WW\, (0j)$ & $0.06 \pm 0.60$ & inclusive & $0.92^{+0.63}_{-0.49}$ & inclusive &    \\
		$WW\, (1j)$ & $2.04^{+1.88}_{-1.30}$ & inclusive & $1.11^{+1.20}_{-0.82}$ & inclusive & \cite{ATLAS:2013wla} \\
		$WW\, (2 j)$ & --- & --- & $1.79^{+0.94}_{-0.75}$ & $(20,80,0)$ & \\
		\midrule
		$ZZ$ & combination  & --- & $1.7^{+0.5}_{-0.4}$ & inclusive& \cite{ATLAS:2013nma} \\
		\midrule
		$\gamma \gamma_{\rm (L)} $ (uc$|$ct) & $0.53^{+1.37}_{-1.44}$ & $(93,7,0)$& $0.86^{+0.67}_{-0.67}$ & $(93.7,6.2,0.2)$& \\
		$\gamma \gamma_{\rm (H)} $ (uc$|$ct) &$0.17^{+1.94}_{-1.91}$ & $(67,31,2)$ & $0.92^{+1.1}_{-0.89}$ & $(79.3,19.2,1.4)$ &  \\
		$\gamma \gamma_{\rm (L)} $ (uc$|$ec) & $2.51^{+1.66}_{-1.69}$& $(93,7,0)$ &$2.51^{+0.84}_{-0.75}$  &  $(93.2,6.6,0.1)$ & \\
		$\gamma \gamma_{\rm (H)} $ (uc$|$ec) & $10.39^{+3.67}_{-3.67}$ & $(65,33,2)$ & $2.69^{+1.31}_{-1.08}$ & $(78.1,20.8,1.1)$& \\
		$\gamma \gamma_{\rm (L)} $ (c$|$ct) & $6.08^{+2.59}_{-2.63}$ & $(93,7,0)$ &  $1.37^{+1.02}_{-0.88}$ &$(93.6,6.2,0.2)$&\\
		$\gamma \gamma_{\rm (H)} $ (c$|$ct) & $-4.40^{+1.80}_{-1.76}$ & $(67,31,2)$ &$1.99^{+1.50}_{-1.22}$  &$(78.9,19.6,1.5)$ & \\
		$\gamma \gamma_{\rm (L)} $ (c$|$ec) & $2.73^{+1.91}_{-2.02}$ & $(93,7,0)$ & $2.21^{+1.13}_{-0.95}$ & $(93.2,6.7,0.1)$&  \\
		$\gamma \gamma_{\rm (H)} $ (c$|$ec) & $-1.63^{+2.88}_{-2.88}$ & $(65,33,2)$ & $1.26^{+1.31}_{-1.22}$ & $(77.7,21.2,1.1)$& \cite{ATLAS:2012goa} \\
		$\gamma \gamma $ (c$|$trans.) & $0.35^{+3.56}_{-3.60}$ & $(89,11,0)$ & $2.80^{+1.64}_{-1.55}$ & $(90.7,9.0,0.2)$& \cite{ATLAS:2013oma} \\
		$\gamma \gamma $ (dijet) & $2.69^{+1.87}_{-1.84}$ & $(23,77,0)$ & --- & --- & \\
		$\gamma \gamma $ ($m_{jj}$ high loose) & --- & --- & $2.76^{+1.73}_{-1.35}$ & $(45,54.9,0.1)$& \\
		$\gamma \gamma $ ($m_{jj}$ high tight) & --- & --- & $1.59^{+0.84}_{-0.62}$ & $(23.8,76.2,0)$& \\
		$\gamma \gamma $ ($m_{jj}$ low) & --- & --- & $0.33^{+1.68}_{-1.46}$ & $(48.1,49.9,1.9)$& \\
		$\gamma \gamma $ ($E_{\rm T}^{\rm miss}$) & --- & --- & $2.98^{+2.70}_{-2.15}$ & $(4.1,83.8,12.1)$& \\
		$\gamma \gamma $ (lepton tag) & --- & --- & $2.69^{+1.95}_{-1.66}$ & $(2.2,79.2,18.6)$& \\
		\bottomrule[1pt]
	\end{tabular}
	\caption{\small ATLAS best fits on signal strength modifier $\mu$ with signal compositions $\zeta^{p}_{i}$ (if provided) for gluon ($g$), vector ($V$), and top ($t$) initiated production \cite{Azatov:2012qz}. If \emph{inclusive} is denoted, the cut efficiencies have been neglected when evaluating $\mu$ from equation \eqref{eq:mu}. If \emph{combination} is denoted, only the $7 \! + \! 8 \, \textrm{TeV}$ combined result is available.}
	\label{tab:ATLAS}
\end{table}

\begin{table}[!ht]
	\centering
	\renewcommand{\arraystretch}{1.1}
	\begin{tabular}{ l | c c | c c | c }
		\toprule[1pt]
		Channel & $\hat \mu$ $(7 \, \textrm{TeV})$ & $\zeta_i^{\rm (g,V,t)}$ (\%) &  $\hat \mu$ $(8 \, \textrm{TeV})$ & $\zeta_i^{\rm (g,V,t)}$(\%) & Refs. \\
		\midrule[1pt]
		$b \bar b$ (VBF) & --- & --- & $0.7^{+1.4}_{-1.4}$ & $(0,100,0)$& \cite{CMS:NOINSP:bbVBF} \\
		$b \bar b$ (VH) & combination & --- & $1.0^{+0.5}_{-0.5}$ & $(0,100,0)$& \cite{CMS:NOINSP:bbVH} \\
		$b \bar b$ ($ttH$)& ${-0.81^{+2.05}_{-1.75} }$ & $(0,30,70)$ & ---  & --- & \cite{CMS:2012ywa}\\
		\midrule
		$\tau \tau$ ($0/1j$)& combination & --- & $0.74^{+0.49}_{-0.52}$ & inclusive &  \\
		$\tau \tau$ (VBF) & combination & ---& $1.38^{+0.61}_{-0.57}$ & $(0,100,0)$ & \cite{CMS:utj} \\
		$\tau \tau$ (VH) & combination & --- & $0.76^{+1.48}_{-1.43}$ & $(0,100,0)$& \\
		\midrule
		$WW\, (0/1 j)$ & combination & --- & $0.76^{+0.21}_{-0.21}$ & inclusive & \\
		$WW\, (2 j)$ & combination & ---  & $-0.05^{+0.73}_{-0.56}$ & $(17,83,0)$ & \cite{CMS:bxa} \\
		$WW$ (VH)  & combination & --- &$-0.31^{+2.24}_{-1.96}$ & $(0,100,0)$& \\
		\midrule
		$ZZ$ (untagged) & combination & --- &  $0.84^{+0.32}_{-0.26}$ & $(95,5,0)$ & \cite{CMS:xwa} \\
		$ZZ$ (dijet tag) & --- & --- &  $1.22^{+0.84}_{-0.57}$ & $(80,20,0)$ & \\
		\midrule
		$\gamma \gamma $ (no tag 0)  & $3.78^{+2.01}_{-1.62}$ & $(61.4, 35.5,3.1)$& $2.12^{+0.92}_{-0.78}$& $(72.9,24.6,2.6)$ & \\
		$\gamma \gamma $ (no tag 1) & $0.15^{+0.99}_{-0.92}$ & $(87.6,11.8,0.5)$ & $-0.03^{+0.71}_{-0.64}$ & $(83.5,15.5,1.0)$ &  \\
		$\gamma \gamma $ (no tag 2) & $-0.05^{+1.21}_{-1.21}$& $(91.3,8.3,0.3)$ & $0.22^{+0.46}_{-0.42}$ & $(91.7,7.9,0.4)$ &  \\
		$\gamma \gamma $ (no tag 3) & $1.38^{+1.66}_{-1.55}$& $(91.3,8.5,0.2)$ & $-0.81^{+0.85}_{-0.42}$ & $(92.5,7.2,0.2)$ &\\
		$\gamma \gamma $ (dijet) & $4.13^{+2.33}_{-1.76}$& $(26.8,73.1,0.0)$ & --- & ---  &  \cite{CMS:ril} \\
		$\gamma \gamma $ (dijet loose) & --- & --- & $0.75^{+1.06}_{-0.99}$ & $(46.8,52.8,0.5)$ & \\
		$\gamma \gamma $ (dijet tight) & --- & --- & $0.22^{+0.71}_{-0.57}$ & $(20.7,79.2,0.1)$ & \\
		$\gamma \gamma $ (MET) & --- & --- & $1.84^{+2.65}_{-2.26}$ & $(0.0,79.3,20.8)$ & \\
		$\gamma \gamma $ (Electron) & --- & --- & $-0.70^{+2.75}_{-1.94}$ &  $(1.1,79.3,19.7)$ & \\
		$\gamma \gamma $ (Muon) & --- & --- & $0.36^{+1.84}_{-1.38}$ &  $(21.1,67.0,11.8)$ & \\
		\bottomrule[1pt]
	\end{tabular}
	\caption{\small CMS best fits on signal strength modifier $\mu$ with signal compositions $\zeta^{p}_{i}$ (if provided) for gluon ($g$), vector ($V$), and top ($t$) initiated production \cite{Azatov:2012qz}. If \emph{inclusive} is denoted, the cut efficiencies have been neglected when evaluating $\mu$ from equation \eqref{eq:mu}. If \emph{combination} is denoted, only the $7 \! + \! 8 \, \textrm{TeV}$ combined result is available.}
	\label{tab:CMS}
\end{table}

\section{EWPO \& precision Higgs: Case B}
\label{sec:higgsewptcaseb}
\begin{figure}[!ht]
	\begin{center}
		\includegraphics[scale=0.6]{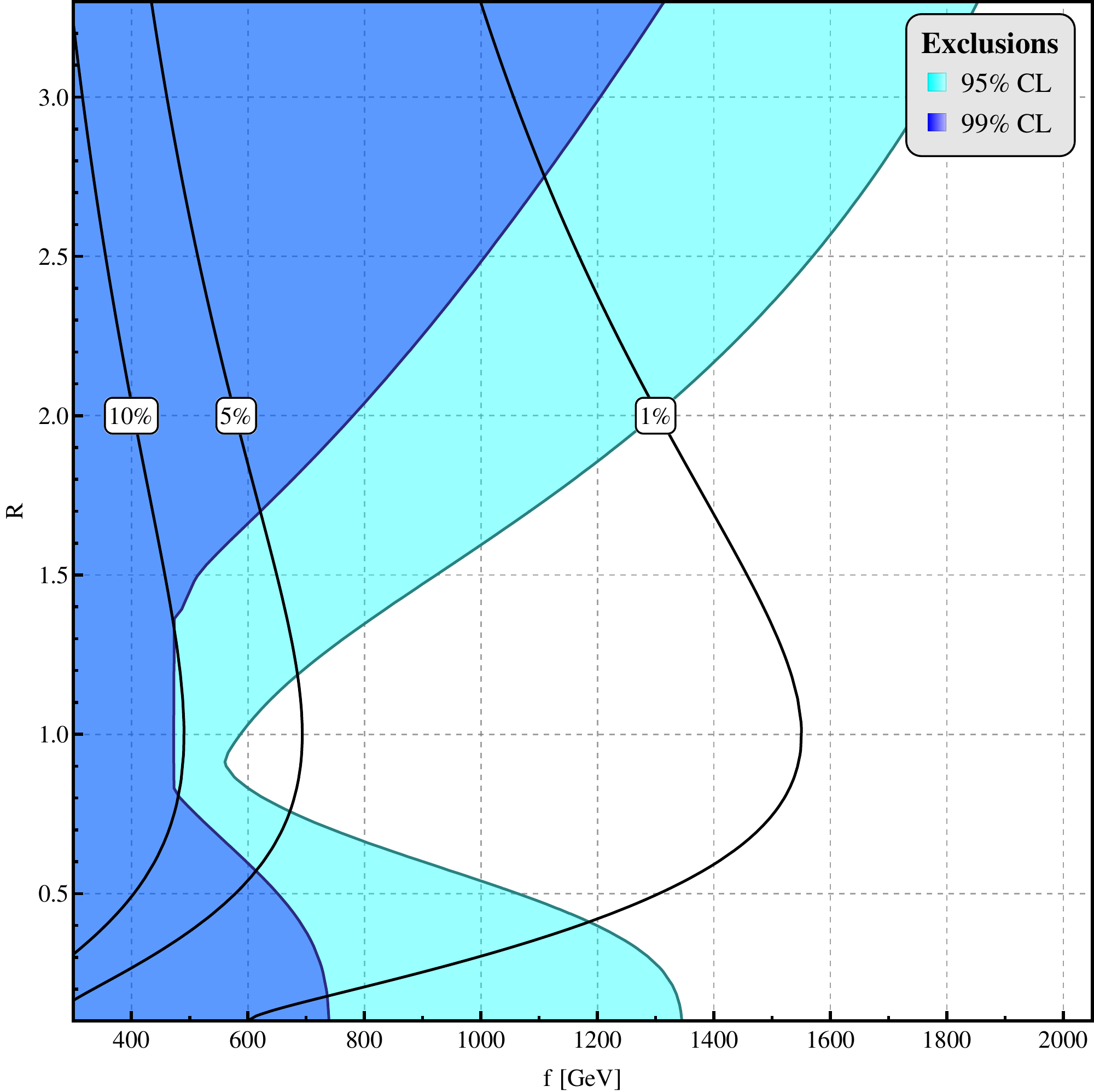} 
		\caption{Excluded parameter space regions at $95\%$ and $99\%$ CL from combination of EWPT and Higgs sector datasets. The thick black lines represent contours of required fine-tuning. Presented here is an alternative description of the Yukawa couplings, known as \emph{Case B}.}
		\label{fig:HiggsEWPTCaseB}
	\end{center}
\end{figure}

In this appendix the combined constraints from EWPT and Higgs searches are presented for a second down-type Yukawa coupling scenario, commonly known as \emph{Case B} \cite{Chen:2006cs}. By combining the $\chi^{2}$ analyses carried out separately for EWPT and the Higgs sector, as plotted in figure \ref{fig:HiggsEWPTCaseB}, the lower bound on the symmetry breaking scale is 
\begin{equation}
	\boxed{f \gtrsim \limitHiggsEWPTCaseB \, \textrm{GeV}} \quad \textrm{at} \quad 95\% \textrm{ CL}.
\end{equation}
The reduced lower bound in \emph{Case B} compared to \emph{Case A} can be explained by the higher suppression in the bottom Yukawa coupling \eqref{eq:caseb}. This in turns yields a higher suppression of the $b \bar{b}$ branching ratio and an enhancement of all other decay rates. This is indeed more aligned with the Higgs results provided by the ATLAS collaboration, where a generic enhancement in the non-fermionic decays of the Higgs is observed. This pattern is not exactly observed in the CMS Higgs results. However, since deviations from the ATLAS results turn out to be dominant in the $\chi^2$ measure, the net result is a weaker exclusion for \emph{Case B}.

\acknowledgments
The authors of this paper are grateful for useful discussions with Masaki Asano, Diptimoy Ghosh, Paolo Gunnellini, Kazuki Sakurai, Andreas Weiler and Lisa Zeune. The authors MT and MdV have been partially supported by the Deutsche Forschungsgemeinschaft within the Collaborative Research Center SFB 676 "Particles, Strings, Early Universe". JRR has been partially supported by the Strategic Alliance ``Physics at the Terascale'' of the Helmholtz Gemeinschaft.

\bibliographystyle{JHEP}
\bibliography{lht}

\end{document}